\documentclass[aps,prx,preprint,onecolumn,citeautoscript,footinbib,
eqsecnum,superscriptaddress]{revtex4-2}  
\usepackage{amsmath,amssymb,bm,bbm} 
\usepackage[compat=1.0.0]{tikz-feynman}
\usepackage{physics}
\usepackage{graphicx}  
\usepackage{float}
\usepackage[caption = false]{subfig}
\usepackage{dsfont} 
\usepackage{color}
\usepackage{xcolor}
\usepackage[papersize={8.5in,11in}]{geometry}
\usepackage[colorlinks=true]{hyperref}  
\usepackage{ulem}
\usepackage[mathscr]{euscript}
\usepackage[section]{placeins}
\usepackage{comment}
\usepackage{tikz}
\hypersetup{
    bookmarks=true,         
    unicode=false,          
    pdftoolbar=true,        
    pdfmenubar=true,        
    pdffitwindow=false,     
    pdfstartview={FitH},    
    pdfsubject={},   
    pdfcreator={},   
    pdfproducer={}, 
    pdfkeywords={} {} {}, 
    pdfnewwindow=true,      
    colorlinks=true,       
    linkcolor=magenta, 
    citecolor=blue,        
    filecolor=magenta,      
    urlcolor=blue           
} 

\usepackage{amsfonts}
\usepackage{upgreek}
\usepackage{slashed}
\usepackage{latexsym}

\newcommand{\beq}{\begin{equation}}
\newcommand{\eeq}{\end{equation}}
\def\bea{\begin{eqnarray}}
\def\eea{\end{eqnarray}}

\newcommand{\bk}{{\bm k}}

\newcommand{\UU}{\operatorname{U}}
\newcommand{\SU}{\operatorname{SU}}
\newcommand{\SO}{\operatorname{SO}}

\newcommand{\vi}{{\boldsymbol{i}}}
\newcommand{\vj}{{\boldsymbol{j}}}
\newcommand{\cCom}{\mathbin{\raisebox{0.5ex}{,}}}
\newcommand{\eq}[1]{\begin{align}#1\end{align}}
\newcommand{\ua}{\uparrow}
\newcommand{\da}{\downarrow}
\newcommand{\nt}{\notag\\}
\newcommand{\id}{\mathds{1}}
\newcommand{\T}{\mathcal{T}}

\definecolor{orange}{rgb}{1,0.5,0}

\newcommand{\vx}{{\boldsymbol{x}}}
\newcommand{\vy}{{\boldsymbol{y}}}

\begin{document}

\title{Deconfined criticality
and a gapless $\mathbb{Z}_2$ spin liquid\\ in the square lattice antiferromagnet}

\author{Henry Shackleton}
\affiliation{Department of Physics, Harvard University, Cambridge MA 02138, USA}

\author{Alex Thomson}
\affiliation{Institute for Quantum Information and Matter, California Institute of Technology, Pasadena, California 91125, USA}
\affiliation{Walter Burke Institute for Theoretical Physics, California Institute of Technology, Pasadena, California 91125, USA}
\affiliation{Department of Physics, California Institute of Technology, Pasadena, California 91125, USA}

\author{Subir Sachdev}
\affiliation{Department of Physics, Harvard University, Cambridge MA 02138, USA}

\date{\today}

\begin{abstract}
The theory for the vanishing of N\'eel order in the spin $S=1/2$ square lattice antiferromagnet has been the focus of attention for many decades. A consensus appears to have emerged in recent numerical studies on the antiferromagnet with first and second neighbor exchange interactions (the $J_1$-$J_2$ model): a gapless spin liquid is present for a narrow window of parameters between the vanishing of the N\'eel order and the onset of a gapped valence bond solid state. We propose a deconfined critical SU(2) gauge theory for a transition  into a stable $\mathbb{Z}_2$ spin liquid with massless Dirac spinon excitations; on the other side the critical point, the SU(2) spin liquid (the `$\pi$-flux' phase) is presumed to be unstable to confinement to the N\'eel phase. We identify a dangerously irrelevant coupling in the critical SU(2) gauge theory, which contributes a logarithm-squared renormalization. This critical theory is also not Lorentz invariant, and weakly breaks the SO(5) symmetry which rotates between the N\'eel and valence bond solid order parameters. We also propose a distinct deconfined critical U(1) gauge theory for a transition into the same gapless $\mathbb{Z}_2$ spin liquid; on the other side of the critical point, the U(1) spin liquid (the `staggered flux' phase) is presumed to be unstable to confinement to the valence bond solid. This critical theory has no dangerously irrelevant coupling, dynamic critical exponent $z \neq 1$, and no SO(5) symmetry. All of these phases and critical points are unified in a SU(2) gauge theory with Higgs fields and fermionic spinons which can naturally realize the observed sequence of phases with increasing $J_2/J_1$: N\'eel, gapless $\mathbb{Z}_2$ spin liquid, and valence bond solid.
\end{abstract}

\maketitle

\tableofcontents

\section{Introduction}
\label{sec:intro}

Antiferromagnetism on the square lattice became a topic of intense study 
soon after the discovery of high temperature superconductivity in the cuprates, and it continues to be a wellspring of interesting experimental and theoretical physics. It was established early on that the insulating antiferromagnet with $S=1/2$ spins on each site, and only nearest neighbor antiferromagnetic exchange interactions ($J_1$) has long-range N\'eel order in its ground state {\it i.e.\/} global SU(2) spin rotation symmetry was broken with the spin expectation value $\langle {\bm S}_{\vi} \rangle = \eta_\vi {\bm N}_0$ where ${\bm S}_\vi$ is the spin operator on site $\vi$, $\eta_\vi = \pm 1$ on the two checkerboard sublattices, and ${\bm N}_0$ is the antiferromagnetic moment. Much attention has since been lavished on the insulating $J_1$-$J_2$ antiferromagnet \cite{IoffeLarkin88,GSH89,Dagotto89,CCL90,RS91,SR91}, which also has a second-neighbor antiferromagnetic exchange interaction $J_2$.
The key questions are the nature of the quantum phases of the model, and of the quantum phase transitions between them, as a function of increasing $J_2/J_1$ after the N\'eel order vanishes at a critical value of $J_2/J_1$. These questions are also the focus of our attention here.

\begin{figure}
\centering
\includegraphics[width=5in]{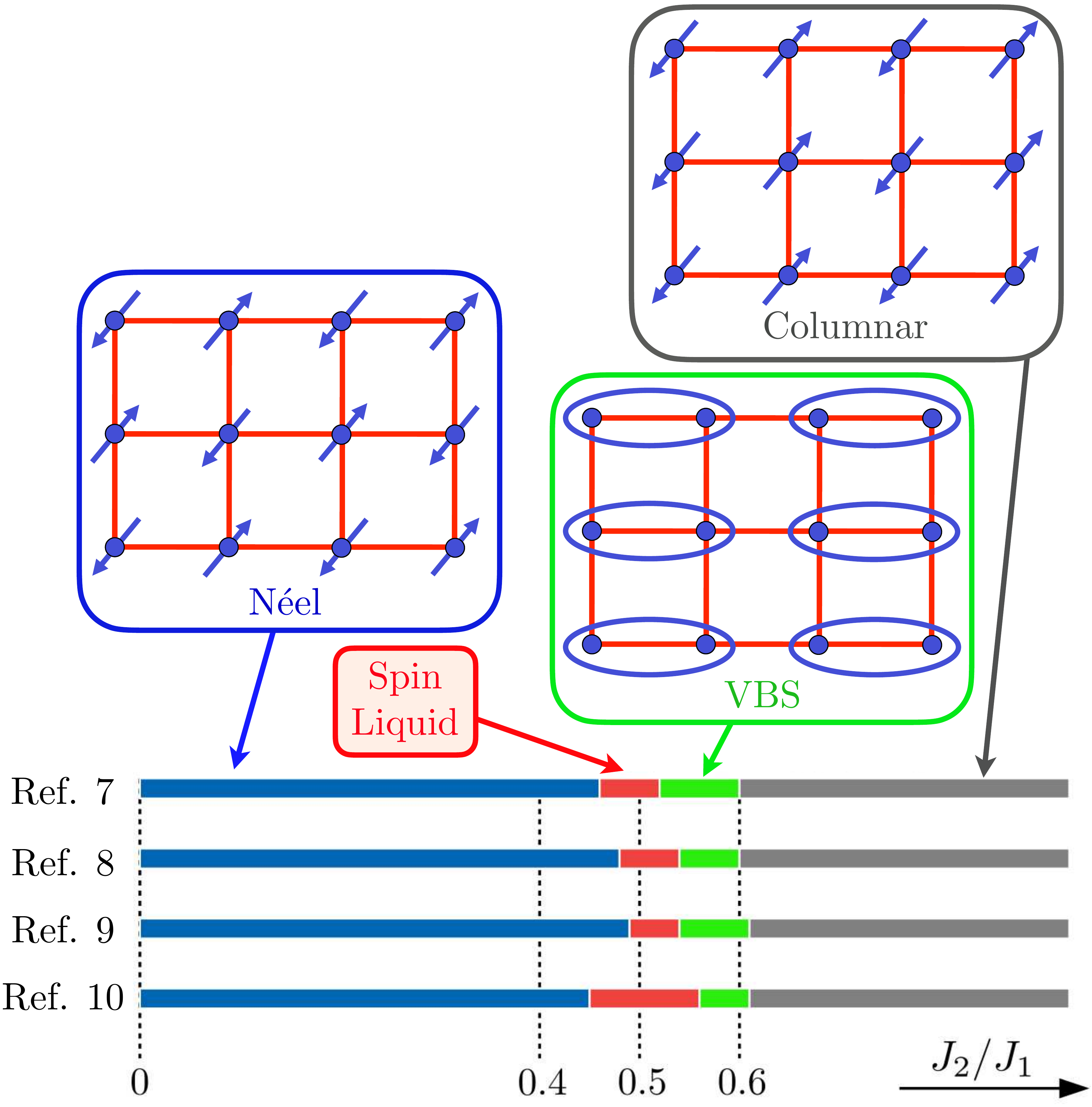}
\caption{Phases of the $S=1/2$ $J_1$-$J_2$ antiferromagnet on the square lattice, from the numerical results of Refs.~\cite{Sandvik18,Becca20,Imada20,Gu20}, all of which agree that the spin liquid is gapless. Each ellipse in the valence bond solid (VBS) represents a singlet pair of electrons. Lower part of figure adapted from Ref.~\cite{BeccaKITP}.}
\label{fig:becca}
\end{figure}
An early proposal \cite{NRSS89,NRSS90,RS91,SR91} was that there was a direct transition from the N\'eel state to a valence bond solid (VBS) (see Fig.~\ref{fig:becca}) which restores spin rotation symmetry but breaks lattice symmetries (followed by a first order transition at larger $J_2/J_1$ to a `columnar' state which breaks spin rotation symmetry, and which we do not address in the present paper). A theory of `deconfined criticality' was developed \cite{OMAV04,senthil1,senthil2} showing that a continuous N\'eel-VBS transition was possible, even though it was not allowed in the Landau-Ginzburg-Wilson framework because distinct symmetries were broken in the two phases. Evidence has since accumulated for the presence of a VBS phase in the $J_1$-$J_2$ model, but the nature of the N\'eel-VBS transition in this model has remained a question of significant debate. However, in the past year, a consensus appears to have emerged \cite{BeccaKITP} among the groups investigating this question by different numerical methods \cite{Sandvik18,Becca20,Imada20,Gu20}, and is summarized in Fig.~\ref{fig:becca}: there is a narrow window with a gapless spin liquid phase between the N\'eel and VBS states. This gapless phase has been identified \cite{Becca01,SenthilIvanov,Becca13,Becca18,Becca20} as a $\mathbb{Z}_2$ spin liquid \cite{RS91,SR91,wen1991,Kitaev1997} with gapless, fermionic, $S=1/2$ spinon excitations with a Dirac-like dispersion \cite{TSMPAF99,WenPSG,SenthilIvanov,SenthilLee05,Kitaev2006}, labeled Z2A$zz13$ in Wen's classification \cite{WenPSG}.

\begin{figure}
    \centering
    \includegraphics[width=5in]{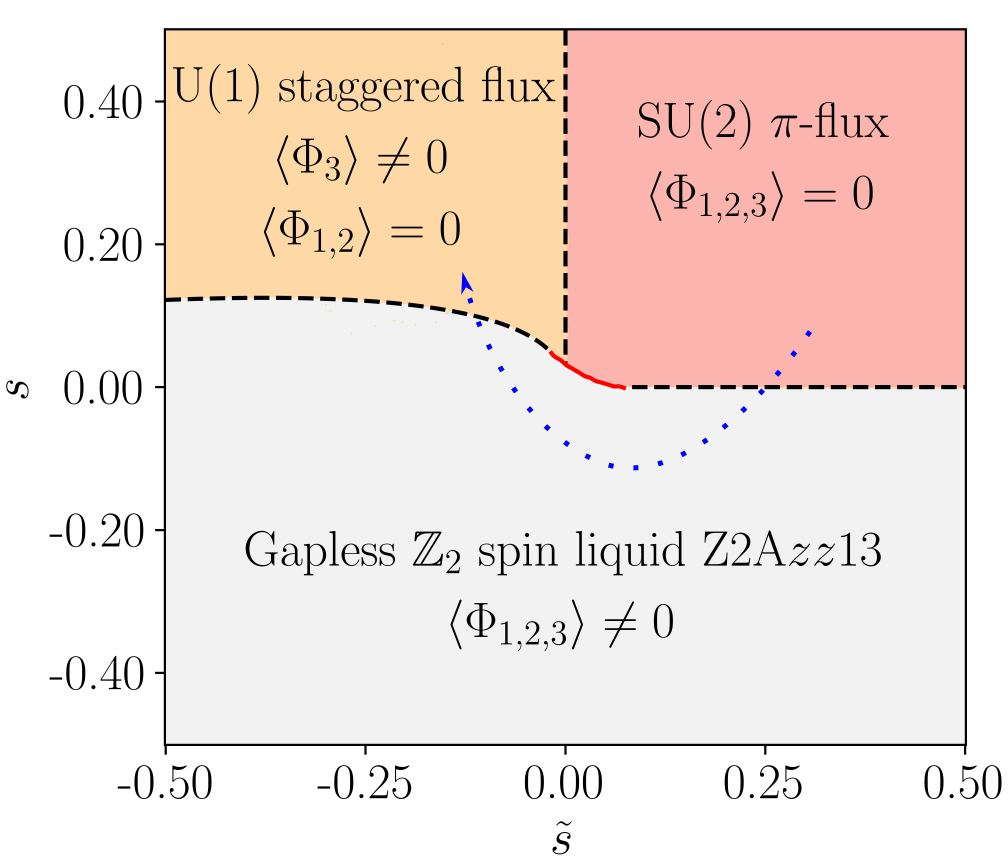}
    \caption{Mean field phase diagram of our low energy theory obtained by minimization of the Higgs potential in Eq.~(\ref{lambda4}). Dashed (solid red) lines indicate second (first) order transitions in mean field theory. We assume the SU(2) $\pi$-flux gauge theory confines to a N\'eel state, the U(1) staggered flux gauge theory confines to a VBS state, except at their deconfined critical boundaries to Wen's stable, gapless $\mathbb{Z}_2$ spin liquid Z2A$zz13$.
    The dotted blue line indicates a possible trajectory of the square lattice antiferromagnet with increasing $J_2/J_1$. However, as we discuss in Section~\ref{sec:conc}, we cannot rule out interchanging the assignments of the confining states between the SU(2) and U(1) spin liquids, in which case the orientation of the blue arrow would be reversed. 
    The critical SU(2) gauge theory has a dangerously irrelevant coupling, 
    but the critical U(1) gauge theory does not.
    The mean-field analysis was performed with $w=u= 1$, $v=-1$, $\tilde{u} = 0.75$, and $v_4 = 0.5$ in Eq.~(\ref{lambda4}). We use the ansatz $\Phi_1^a = c_1 \delta_{ax}$, $\Phi_2^a = c_1 \delta_{ay}$, and $\Phi_3^a = c_2 \delta_{az}$, so the terms in $V(\Phi)$ proportional to $v_1, v_3$ are automatically zero. }
    \label{fig:mfPhaseDiagram}
\end{figure}

The starting point of our analysis is the fermionic spinon dual \cite{Wang17,Thomson17,Song1,Song2} of the bosonic spinon CP$^1$ model used earlier \cite{NRSS89,NRSS90,senthil1,senthil2} to describe the N\'eel-VBS transition. This fermionic dual is a relativistic SU(2) gauge theory  of 2 flavors of 2-component, massless Dirac fermions carrying fundamental gauge charges: this formulation is preferred over the bosonic spinons because the massless Dirac fermions connect naturally to the gapless fermionic spinons in the $\mathbb{Z}_2$  spin liquid. Recent studies \cite{Wang19,Nahum19,Assaad21,He20} have indicated that the 2 fermion flavor SU(2) gauge theory does not ultimately describe a conformal field theory needed for N\'eel-VBS criticality, but exhibits a `pseudocriticality' associated with a proximate fixed point at complex coupling \cite{Wang17,Gorbenko:2018ncu,Gorbenko:2018dtm,Ma:2018euv}. Ref.~\cite{Thomson17} used connections to bosonic spinon theories to argue that the 2 fermion flavor SU(2) gauge theory was ultimately unstable to confinement and symmetry breaking leading to the appearance of N\'eel order. We assume this is the case, and we can then describe the transition to the $\mathbb{Z}_2$ spin liquid by the condensation of Higgs fields which break the SU(2) gauge symmetry down to $\mathbb{Z}_2$: see Fig.~\ref{fig:mfPhaseDiagram}. The N\'eel-$\mathbb{Z}_2$ spin liquid transition is a confinement-Higgs transition, and the critical theory is proposed to be a 2-flavor SU(2) gauge theory with critical Higgs fields \cite{Thomson17}. 
We note that a similar critical theory was proposed in Ref.~\cite{Gazit18} for a continuous transition from the N\'eel state to a different gapless state with a $\mathbb{Z}_2$ gauge field (the `orthogonal semi-metal'), and this scenario was supported there by quantum Monte Carlo simulations. Evidently, it is possible that critical Higgs fields can stabilize a scale-invariant critical point of the 2-flavor SU(2) gauge theory at the boundary of a Higgs phase where the SU(2) gauge symmetry is broken down to $\mathbb{Z}_2$.

As we will see below, an important difference between our critical Higgs SU(2) gauge theory and that of Ref.~\cite{Gazit18} is that our theory does not preserve Lorentz invariance. The Lorentz symmetry is broken by the Yukawa couplings between the Higgs fields and fermions. The Yukawa couplings also do not preserve the SO(5) flavor symmetry of the SU(2) gauge theory with only fermionic matter; this symmetry rotates between the N\'eel and VBS states. Both these features have important consequences for the N\'eel-$\mathbb{Z}_2$ spin liquid critical point, and lead to predictions described below which can tested by numerical studies. 

In earlier work, Ran and Wen \cite{RanWen06,YingRanThesis}
had considered the 2-flavor SU(2) gauge theory as the description of an extended gapless phase on the square lattice---also called the $\pi$-flux phase \cite{Affleck_1988}. They proposed a theory for a transition from the $\pi$-flux phase to the Z2A$zz13$ spin liquid by the condensation of a pair of adjoint Higgs fields, which we denote $\vec{\Phi}_{1,2}$ (the vector symbol implies gauge SU(2) adjoint index). In light of our arguments above on the confining instability of the $\pi$-flux phase to the N\'eel state, the critical Higgs theory of Ran and Wen \cite{RanWen06,YingRanThesis} can serve as the deconfined critical theory for the N\'eel to Z2A$zz13$ spin liquid transition. However, as we shall see in Section~\ref{sec:rg}, additional `dangerously irrelevant' terms are needed to fully define the critical theory in a $1/N_f$ expansion, and these contribute a logarithm-squared renormalization.

The mean-field phase diagram of the SU(2) gauge theory with adjoint Higgs fields describing the $\pi$-flux to Z2A$zz13$ transition turns out to naturally acquire an additional phase, as explained in Section~\ref{sec:MajoranaHiggs}---this is the U(1) staggered flux spin liquid \cite{Affleck_1988}. We show that the adjoint Higgs field 
\beq
\vec{\Phi}_3 \sim \vec{\Phi}_1 \times \vec{\Phi}_2 \label{Phicross}
\eeq
(see Eq.~\eqref{Phi321}) is precisely that required to go from the SU(2) $\pi$-flux phase to the U(1) staggered flux phase. Specifically, starting from the $\pi$-flux phase, if both $\vec{\Phi}_{1,2}$ condense with $\langle \vec{\Phi}_1 \rangle \times \langle \vec{\Phi}_2 \rangle \neq 0$, we obtain the gapless $\mathbb{Z}_2$ spin liquid (the simultaneous condensation of $\vec{\Phi}_1$ and $\vec{\Phi}_2$ does not require fine tuning because of symmetry constraints that we will describe). On the other hand, Eq.~(\ref{Phicross}) implies that if only the composite field $\vec{\Phi}_1 \times \vec{\Phi}_2$ condenses, but the individual fields $\vec{\Phi}_{1,2}$ do not, then the $\pi$-flux phase turns into the U(1) staggered flux phase. Speaking imprecisely, starting from the parent $\pi$-flux phase, the Higgs condensate for the gapless $\mathbb{Z}_2$ spin liquid is the `square root' of the Higgs condensate for the staggered flux phase. (Let us also note that Song {\it et al.} \cite{Song1} proposed that a trivial monopole would drive the the staggered flux state into the $\pi$-flux state: so the Higgs field $\vec{\Phi}_3$ can be viewed as a `dual' description of the trivial monopole, and induces a transition in the opposite direction. Four-fermion terms have also been proposed as a route to reducing the emergent symmetry of the staggered flux state to that of the $\pi$-flux state \cite{CXSS08}.)
The phase diagram of the Higgs fields $\vec{\Phi}_{1,2,3}$ is computed in Section~\ref{sec:MajoranaHiggs} and shown in Fig.~\ref{fig:mfPhaseDiagram}. We propose here that the transition from the gapless $\mathbb{Z}_2$ spin liquid to the VBS state is described by the deconfined critical theory appearing at the onset of the U(1) spin liquid. Other works \cite{SenthilLee05,Song1,Song2} have discussed the possible instability of this U(1) spin liquid  to either N\'eel or VBS order via monopole proliferation. The critical U(1) gauge theory is described briefly in Section~\ref{sec:sf}, where we show that it does {\it not\/} contain the dangerously irrelevant terms found in the critical SU(2) theory. 

We will review the derivation of the Ran-Wen theory, and discuss its symmetry properties in some detail in Section~\ref{sec:gaplessz2} and Appendix~\ref{app:psg}. A continuum SU(2) gauge theory coupled to 3 adjoint Higgs fields and gapless Majorana fermions will be obtained in Section~\ref{sec:Higgs}. The critical SU(2) gauge theory for the onset of the gapless $\mathbb{Z}_2$ spin liquid phase from the $\pi$-flux phase will be presented in Section~\ref{sec:rg}, along with an analysis of its properties in a $1/N_f$ expansion.
The critical U(1) gauge theory for the onset of the same gapless $\mathbb{Z}_2$ spin liquid from the staggered flux phase appears in Section~\ref{sec:sf}.

\section{\texorpdfstring{Gapless $\mathbb{Z}_2$ spin liquid}{Gapless Z2 spin liquid}}
\label{sec:gaplessz2}

The fermionic spinon theory of $\mathbb{Z}_2$ spin liquids proceeds by re-expressing the spin operators in terms of spinons $f_{\vi \alpha}$, $\alpha = \uparrow, \downarrow$ at site $\vi=(i_x,i_y)$ of the square lattice using
\beq
{\bm S}_\vi = \frac{1}{2} \sum_{\alpha,\beta} f_{\vi \alpha}^\dagger {\bm \sigma}_{\alpha\beta} f_{\vi \beta}\,.
\eeq
We write down a Bogoliubov Hamiltonian for the $f_{\vi \alpha}$ to obtain a $\mathbb{Z}_2$ spin liquid. Following Wen's notation \cite{WenPSG}, we introduce the Nambu spinor
\beq
\psi_\vi = \left( \begin{array}{c} f_{\vi\uparrow} \\ f_{\vi\downarrow}^\dagger \end{array} \right)\,,
\label{eq:Nambu}
\eeq
resulting in the Bogoliubov Hamiltonian
\beq
H = - \sum_{\vi\vj } \psi_\vi^\dagger u_{\vi\vj } \psi_j\,. \label{bogoliubov}
\eeq
Here, 
\beq
u_{\vi\vj } =  i u_{\vi\vj }^0 + u_{\vi\vj }^x \tau^x + u_{\vi\vj }^y \tau^y+ u_{\vi\vj }^z \tau^z\,,
\eeq
with $\tau^a$ Pauli matrices acting on the Nambu indices of $\psi_\vi$. Invariance under global SU(2)$_s$ spin rotation requires that the  $u^\mu_{\vi\vj }$ are all real numbers obeying
\beq
u_{\vj \vi}^0 = - u_{\vi\vj }^0, \quad u_{\vj \vi}^x =  u_{\vi\vj }^x, \quad u_{\vj \vi}^y =  u_{\vi\vj }^y, \quad u_{\vj \vi}^z =  u_{\vi\vj }^z\,.
\label{eq:usu2}
\eeq
This fermionic spinon representation has a SU(2)$_g$ gauge symmetry, under which
\eq{\label{eqn:LatticeGaugeAction0}
\mathrm{SU}(2)_g : \psi_\vi \to U_{g,\vi} \psi_\vi.
}
and a corresponding transformation for $u_{\vi\vj}$.

\begin{figure}
\centering
\includegraphics[width=2in]{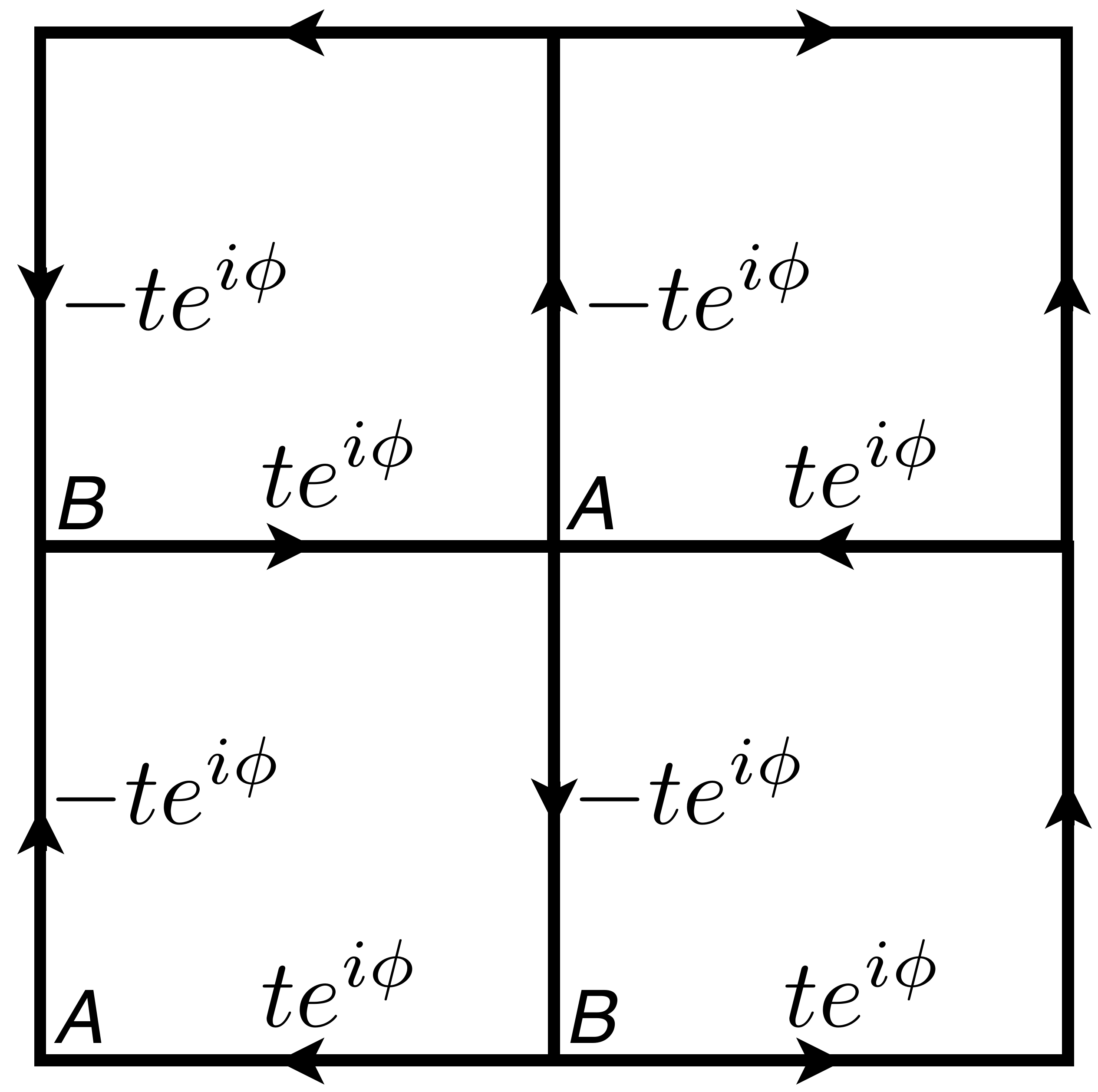}
\caption{Nearest-neighbor fermionic spinon hopping showing the $A$ ($i_x + i_y$ even) and $B$ ($i_x + i_y$ odd) sublattices.}
\label{fig:sf}
\end{figure}
We will provide 3 different ansatzes for the $u_{\vi\vj}$ in the Z2A$zz13$ spin liquid, each suited for different purposes. The 3 ansatzes are, of course, related to each other by SU(2)$_g$ gauge transformations.
Wen's ansatz for the Z2A$zz13$ spin liquid is given in Appendix~\ref{app:psg}, where the continuum Lagrangian describing the different spin liquid phases is deduced from symmetry fractionalization considerations. In the main text, we obtain the continuum theory directly from the lattice model, for which the ansatz given in Eq.~\eqref{su2ansatz} will be most useful. To derive this ansatz, we first describe the Z2A$zz13$ spin liquid by starting from the familiar staggered flux phase with U(1) gauge symmetry \cite{Affleck_1988}, and perturbing it with $d_{xy}$ pairing. Explicitly, the ansatz is 
\bea
\tilde{u}_{\vi,\vi+\hat{x}}  &=&  \left( \begin{array}{cc} te^{-i \phi} & 0 \\
0 & -t e^{i \phi} \end{array} \right), \quad \mbox{$i_x + i_y$ even} \nonumber \\
\tilde{u}_{\vi,\vi+\hat{x}}  &=&  \left( \begin{array}{cc} te^{i \phi} & 0 \\
0 & -t e^{-i \phi} \end{array} \right), \quad \mbox{$i_x + i_y$ odd} \nonumber \\
\tilde{u}_{\vi,\vi+\hat{y}}  &=&   \left( \begin{array}{cc} -te^{i \phi} & 0 \\
0 & t e^{-i \phi} \end{array} \right), \quad \mbox{$i_x + i_y$ even} \nonumber \\
\tilde{u}_{\vi,\vi+\hat{y}}  &=&   \left( \begin{array}{cc} -te^{-i \phi} & 0 \\
0 & t e^{i \phi} \end{array} \right), \quad \mbox{$i_x + i_y$ odd} \nonumber \\
\tilde{u}_{\vi,\vi+\hat{x}+\hat{y}}  &=&  \left( \begin{array}{cc} 0 & -(\gamma_1 - i \gamma_2)  \\
-(\gamma_1 + i \gamma_2)  & 0 \end{array} \right)  \nonumber \\
\tilde{u}_{\vi,\vi-\hat{x}+\hat{y}}  &=&  \left( \begin{array}{cc} 0 & (\gamma_1 - i \gamma_2) \\
(\gamma_1 + i \gamma_2) & 0 \end{array} \right) \label{sfansatz}.
\eea
The first four terms in \eqref{sfansatz} represent the fermion hopping, which is sketched in Fig.~\ref{fig:sf}, and the last 2 terms are the $d_{xy}$ pairing. 
With this ansatz, three distinct spin liquids may be described depending on the choice of parameters. These spin liquids are shown in Fig.~\ref{fig:mfPhaseDiagram}, and we list them below:
\begin{itemize}
    \item The $\pi$-flux phase with SU(2) gauge symmetry corresponds to $\phi=\pi/4$, and no fermion pairing $\gamma_{1,2}=0$.
    \item 
    The `staggered flux' U(1) spin liquid is obtained for general $\phi$, and no fermion pairing $\gamma_{1,2}=0$. The U(1) gauge field corresponds to a nearly spatially uniform modulation in the phases of the fermion hopping terms.
\item The Z2A$zz13$ spin liquid is obtained when the $d_{xy}$ pairing $\gamma_1+i \gamma_2$ is present, and breaks the U(1) down to $\mathbb{Z}_2$. 
\end{itemize}
Note that we have $d_{xy}$ pairing in the $\mathbb{Z}_2$ spin liquid only, with opposite signs on the two sublattices.

In momentum space, we choose the $A$ and $B$ checkerboard sublattices as the basis sites (shown in Fig.~\ref{fig:sf}), and the Hamiltonian acting on $\left(f_{A,\bk\uparrow} , f_{B,\bk\uparrow}, f_{A,-\bk,\downarrow}^\dagger,f_{B,-\bk,\downarrow}^\dagger \right)^T$ in the gauge of Eq.~(\ref{sfansatz}) is 
\beq
H = \left(
\begin{array}{cccc}
0 & C_\bk & D_\bk & 0 \\
C_\bk^\ast & 0 & 0 & D_\bk \\
D_\bk^\ast & 0 & 0 & -C_\bk^\ast \\
0 & D_\bk^\ast & -C_\bk & 0 
\end{array}
\right) \label{e1}
\eeq
where
\beq
C_\bk = -2 t (e^{-i \phi} \cos(k_x) - e^{i \phi} \cos (k_y)) \quad, \quad D_\bk = 4 (\gamma_1 - i \gamma_2) \sin (k_x) \sin(ky) \,.
\eeq
The eigenvalues of (\ref{e1}) are
\beq
\varepsilon_{\bk} = \pm \left(\left[ \mbox{Re} (C_\bk)\right]^2 + \left[\mbox{Im} (C_\bk) \pm |D_\bk| \right]^2 \right)^{1/2} \label{e2a}
\eeq
and these co-incide with those obtained in Wen's gauge in (\ref{e2}). Note that the dispersion depends only on $|\gamma_1 + i \gamma_2|$, and not on $\gamma_{1,2}$ separately. This is natural in the staggered flux gauge, where U(1) the gauge transformation acts simply as $f_{\vi \alpha} \rightarrow f_{\vi \alpha} e^{i \phi_\vi}$, and so the $d_{xy}$ pairing acts like a charge 2 Higgs field: a simple identification of the charge 2 Higgs field is the advantage of the present gauge.
\begin{figure}
\centering
\includegraphics[width=4.5in]{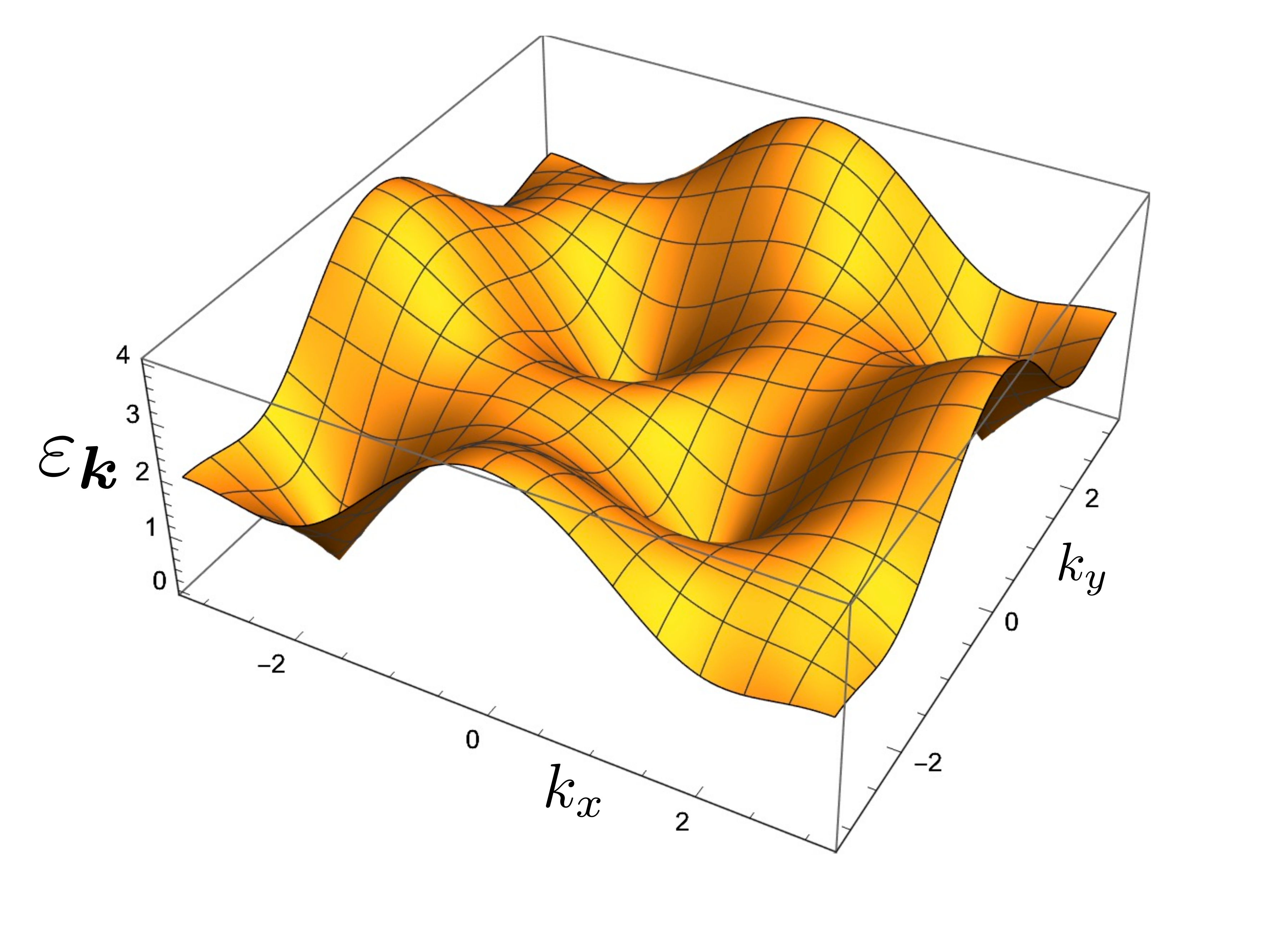}
\caption{Plot of the dispersion, $\varepsilon_\bk$, of the fermionic spinons of the $\mathbb{Z}_2$ spin liquid Z2A$zz13$. The eigenvalues of the spinon Hamiltonian are $\pm \varepsilon_{\bk}$. All gauge invariant observables are invariant under the square lattice space group, although the spinon dispersion is not. The plot is of Eq.~(\ref{e2a}) for $t=1.118$, $\phi=0.464$, $\gamma_1 = 0.5$, $\gamma_2=0$.}
\label{fig:disp}
\end{figure}
This dispersion is plotted in Fig.~(\ref{fig:disp}). 
The staggered flux phase has Dirac nodal points at $(\pm \pi/2, \pm \pi/2)$. Introducing $d_{xy}$ pairing does not gap these nodal points, but moves them away from these high symmetry points. 
Although the dispersion does not have full square lattice symmetry, all gauge-invariant observables do, and this is verified by the analysis in Appendix~\ref{app:psg}.

\subsection{Majorana gauge}
\label{sec:Majorana}

For the remainder of the analysis in the body of the paper we map (\ref{sfansatz}) onto the gauge used by Wang {\it et al.} \cite{Wang17} for the $\pi$-flux phase, which is convenient for eventual representation in Majorana fermions and making the gauge and spin rotation symmetries manifest.
In this gauge, the ansatz of the Z2A$zz13$ spin liquid (which is gauge equivalent to Eq.~(\ref{sfansatz})) is
\bea
\bar{u}_{\vi,\vi+\hat{{\bm x}}}  &=& \left( \begin{array}{cc} ite^{-4 i \phi} & 0 \\
0 & it e^{4 i \phi} \end{array} \right) \quad, \quad \mbox{$i_x + i_y$ even} \nonumber \\
\bar{u}_{\vi,\vi+\hat{{\bm x}}}  &=&  \left( \begin{array}{cc} ite^{4 i \phi} & 0 \\
0 & it e^{-4 i \phi} \end{array} \right) \quad, \quad \mbox{$i_x + i_y$ odd}  \nonumber \\
\bar{u}_{\vi,\vi+\hat{{\bm y}}}  &=&  (-1)^{i_x} \left( \begin{array}{cc} te^{2 i \phi} & 0 \\
0 & -t e^{-2 i \phi} \end{array} \right) \quad, \quad \mbox{$i_x + i_y$ odd}  \nonumber \\
\bar{u}_{\vi,\vi+\hat{{\bm y}}}  &=&  (-1)^{i_x} \left( \begin{array}{cc} -te^{-2 i \phi} & 0 \\
0 & t e^{2 i \phi} \end{array} \right) \quad, \quad \mbox{$i_x + i_y$ even}  \nonumber \\
\bar{u}_{\vi,\vi+\hat{{\bm x}}+\hat{{\bm y}}} =\bar{u}_{\vi,\vi-\hat{{\bm x}}+\hat{{\bm y}}} &=&  \left( \begin{array}{cc} 0 & ( \gamma_1 -  i\gamma_2) e^{-2i\phi} \\
( \gamma_1 + i\gamma_2)e^{2i\phi} & 0 \end{array} \right)   \quad, \quad \mbox{$i_x + i_y$ even} \nonumber \\
\bar{u}_{\vi,\vi+\hat{{\bm x}}+\hat{{\bm y}}} =\bar{u}_{\vi,\vi-\hat{{\bm x}}+\hat{{\bm y}}} &=&   \left( \begin{array}{cc} 0 & (-\gamma_1 +i \gamma_2) e^{4i \phi}  \\
(-\gamma_1 - i\gamma_2) e^{-4 i \phi} & 0 \end{array} \right)   \quad, \quad \mbox{$i_x + i_y$ odd} 
\label{su2ansatz}
\eea
As in the previous gauge, the $\pi$-flux phase is obtained when $\phi=\pi/4$ while the staggered-flux phase corresponds to general $\phi$. 

\section{Continuum theory for Higgs transition from SU(2) to $\mathbb{Z}_2$.}
\label{sec:Higgs}

\subsection{$\pi$-flux state with SO(5) symmetry}
\label{sec:pifluxso5}

We begin by working out the continuum SU(2) gauge theory with the 2-flavor massless Dirac fermion from the mean-field ansatz for the $\pi$-flux phase, using the Majorana gauge given in Eq.~\eqref{su2ansatz}. 

In this gauge, we replace the Nambu spinor in Eq.~(\ref{eq:Nambu}) by the matrix operator 
\eq{
\mathcal{X}_\vi&=
\begin{pmatrix}
f_{\vi\uparrow} &   -f^\dag_{\vi\da} \\
f_{\vi\da}  &   f^\dag_{\vi\ua} 
\end{pmatrix}
}
The spinon $\SU(2)$ gauge symmetry of Eq.~(\ref{eqn:LatticeGaugeAction0}) now acts on $\mathcal{X}_\vi$ as
\eq{\label{eqn:LatticeGaugeAction}
\mathrm{SU}(2)_g : \mathcal{X}_\vi \to \mathcal{X}_\vi U_{g,\vi}^\dag.
}
The physical spin symmetry acts on $\mathcal{X}_\vi$ on the left:
\eq{
\mathrm{SU}(2)_s:   \mathcal{X}_\vi \to U_{s}\mathcal{X}_\vi.
}
We write the Bogoliubov Hamiltonian Eq.~(\ref{bogoliubov}) as
    \begin{equation}
      \begin{aligned}
        H_{MF} &= \sum_{\langle i j \rangle} \left[ i \alpha_{\vi\vj } \Tr\left( \mathcal{X}_\vi^\dagger \mathcal{X}_j \right) + \beta_{\vi\vj }^a \Tr\left( \sigma^a \mathcal{X}_\vi^\dagger \mathcal{X}_j \right) + i \gamma_{\vi\vj } \Tr\left( \sigma^a \mathcal{X}_\vi^\dagger \sigma^a \mathcal{X}_j \right) \right]\,.
        \label{eq:initialMFT}
      \end{aligned}
    \end{equation}
The correspondence with the notation in Eq.~(\ref{bogoliubov}) is 
\begin{equation}\label{eqn:MajaronaDiracLatticeReln}
    u_{\vi\vj } = i\alpha_{\vi \vj} \tau^0 + \beta^a_{\vi \vj} \tau^a\,.
\end{equation} 
The additional $\gamma_{\vi\vj }$ hoppings involve projective realizations of the spin rotation symmetry, and will not be relevant. 
    The degrees of freedom in this Hamiltonian can be represented by four Majorana fermions, 
\beq
\mathcal{X}_\vi = \frac{1}{\sqrt{2}} \left( \chi_0 + i \chi_a \sigma^a \right).
\label{eq:Xchi}
\eeq

The $\SU(2)$-invariant $\pi$-flux state comes from the hoppings $\beta^a = 0$ and
    \begin{equation}
      \begin{aligned}
        \alpha_{\vi\vj } = - \alpha_{\vj \vi} \quad\quad \alpha_{\vi + \hat{{\bm x}}, \vi} = 2t \quad\quad \alpha_{\vi+\hat{{\bm y}}, \vi} = (-1)^{i_x} 2t\,.
      \end{aligned}
    \end{equation}
The low-energy behavior of this mean-field ansatz is described by an $\SU(2)$ gauge theory with an emergent $\SO(5)$ symmetry. To work out the dispersion relation of this Hamiltonian, we increase our unit cell by one lattice site in the $x$ direction and so $\chi$ acquires an additional sublattice index $m=A,B$. Note that this unit cell differs slightly from the one used in the staggered flux gauge.
In momentum space, we then have 
    \begin{equation}
      \begin{aligned}
        H &= \sum_{k} \chi_{-k}^\text{T} H(k) \chi_{k}\,,
        \\
        H(k) &= -2t\left[ \sin(k_y) \rho^z + \sin(k_x) \rho^x \right]\,.
      \end{aligned}
      \label{eq:Hk}
    \end{equation}
     $\rho^i$ are Pauli operators acting on the sublattice space, $m=A,B$. This Hamiltonian is diagonal in the $0,a$ indices in Eq.~(\ref{eq:Xchi}), and the gauge was chosen to have this feature. 
     The Hamiltonian in Eq.~\eqref{eq:Hk} has Dirac points at $k_y = 0, \pi$, $k_x = 0$. Labelling these Dirac points by another index $v=1,2$,  and expanding around these two points, we decompose our Majorana operator as
    \begin{equation}
      \begin{aligned}
        \chi_{m, \vi} \sim \rho^x \chi_{m,v=1}(x) + (-1)^{\vi_y} \chi_{m, v=2}(x)\,.
        \label{eq:decomposition}
      \end{aligned}
    \end{equation}
    With this, the Hamiltonian reduces to
    \begin{equation}
      \begin{aligned}
        H \approx 2it\sum_{v=1,2} \chi_{ v}^T \left( \rho^x \partial_x - \rho^z \partial_y \right) \chi_{ v}\,,
      \end{aligned}
    \end{equation}
    with the sublattice and $0,a$ indices implicit.
    This gives the continuum Lagrangian
    \begin{equation}
      \begin{aligned}
        \mathcal{L}_{MF} = 2 it \, \bar{\chi}_{v} \gamma^\mu \partial_\mu \chi_{v}
      \end{aligned}
    \end{equation}
    where $\gamma^0 = \rho^y$, $\gamma^x = i \rho^z$, $\gamma^y = i \rho^x$, and $\bar{\chi} \equiv \chi^T \gamma^0$. 
    Here we have chosen to express $\mathcal{L}_{MF}$ in the Minkowski metric $(+,-,-)$; we ultimately move to the Euclidean metric below to perform calculations. 

    We now define the $4 \times 2$ matrix operator
    \begin{equation}
      \begin{aligned}
        X_{\alpha, v; \beta} = \frac{1}{\sqrt{2}} \left( \chi_{0, v} \delta_{\alpha \beta} + i \chi_{a, v} \sigma^a_{\alpha\beta} \right)
        \label{eq:matrixMajorana}
      \end{aligned}
    \end{equation}
    and $\bar{X} = X^\dagger \gamma^0$, where the sublattice/Dirac index $m$ is left implicit. This
    lets us write our Lagrangian as
    \begin{equation}
      \begin{aligned}
        \mathcal{L}_{MF} = i \Tr \left( \bar{X} \gamma^\mu \partial_\mu X \right)\,,
      \end{aligned}
    \end{equation}
    where we set $t=1/2$ from now on.
In this form, the Hamiltonian describes 8 massless Majorana fermions
(these are 2-component `relativistic' Majorana fermions with an additional sublattice index). The SU(2) gauge symmetry acts on the right index ($\beta$ in Eq.~\eqref{eq:matrixMajorana}) of $X$, and the gradient in $\mathcal{L}_{MF}$ must be replaced by the appropriate covariant gradient when the gauge field is included.
Global spin rotations act of the left index ($\alpha$ in Eq.~\eqref{eq:matrixMajorana}) of $X$, and global valley rotations act of the $v$ index. These global rotations combine to yield an emergent, low energy  Sp(4)$/\mathbb{Z}_2 \equiv$ SO(5) global symmetry in the $\pi$-flux phase \cite{Wang17,RanWen06}.

In the following subsections, we derive the continuum form of the perturbations given in Eq.~(\ref{su2ansatz}), which break the $\pi$-flux state down to either the staggered flux state or the Z2A$zz13$ spin liquid. We do so by rewriting these perturbations in terms of the low-energy modes given in Eq.~(\ref{eq:decomposition}) and keeping only the lowest order gradient terms. These perturbations are coupled to adjoint Higgs fields, and the transition of the $\pi$-flux state to either the staggered flux state or Z2A$zz13$ spin liquid is obtained by condensing the corresponding Higgs fields. An alternative derivation of these continuum perturbations based on symmetry fractionalization is provided in Appendix~\ref{app:psg}, and agrees with the following analysis.

\subsection{From $\pi$-flux to staggered flux}
\label{sec:pitostag}
\begin{figure}
    \centering
\includegraphics[width=2in]{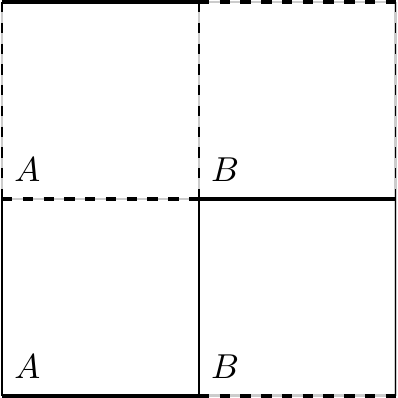}
    \hspace{2cm}
    \includegraphics[width=2in]{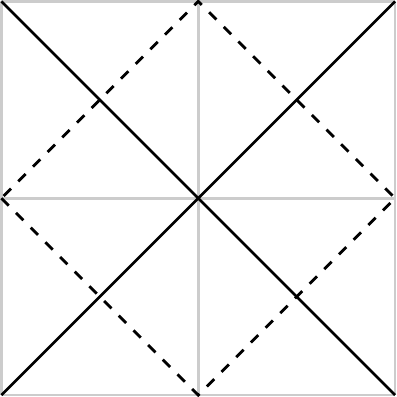}
    \caption{Shown are the leading-order perturbations that away from the $\SU(2)$ $\pi$-flux state, in the Majorana gauge given by Eq.~\ref{su2ansatz}. Note that the unit cell, with sublattice sites $A$ and $B$, differs from the gauge illustrated in Fig.~\ref{fig:sf}. (Left) The perturbation that shifts the $\pi$-flux state to the staggered flux state, whit hoppings proportional to $\tau^z$. Thickness of the line denotes strength (weaker in the $y$-direction) and solid/dashed indicates positive/negative sign. (Right) The $d_{xy}$ pairing that breaks the $\UU(1)$ gauge symmetry to $\mathbb{Z}_2$, with pairing $\gamma_1 \tau^y - \gamma_2 \tau^x$ on solid lines and $\gamma_1 \tau^x + \gamma_2 \tau^y$ on dashed lines.}
    \label{fig:piFluxPerturbations}
\end{figure}
We obtain the continuum version of the perturbations to the staggered flux phase by expanding the mean field parameters $\bar{u}_{\vi\vj}$ defined in Eq.~\eqref{su2ansatz} in powers of $\phi=\pi/4 + \delta\phi$. We subsequently employ Eq.~\eqref{eqn:MajaronaDiracLatticeReln}, which in turn yields additional hopping parameters to the Hamiltonian of the form
\begin{align}
\beta^z_{\vi,\vi+\hat{\vx}}&=-4\delta\phi(-1)^{i_x + i_y},
&
\beta^z_{\vi,\vi+\hat{\vy}}&=2\delta\phi(-1)^{i_y}.
\end{align}
These terms are illustrated in Fig.~\ref{fig:piFluxPerturbations}.
If we look at the components of the Majorana fermions (as defined in Eq.~\eqref{eq:Xchi}, with $(0,a)$, $a=x,y,z$), we see that these new terms introduce hopping between the $0\leftrightarrow z$ and $x\leftrightarrow y$ Majorana fermions.
For simplicity, we focus on the $0 \leftrightarrow z$ hoppings, as the $x \leftrightarrow y$ hoppings will be identical. We start with the hoppings in the $x$-direction, expand our Majorana operators in terms of low-energy modes, and keep only the lowest-order gradient terms. As in Eq.~\eqref{eq:matrixMajorana}, the two indices on $\chi$ correspond to $(0,x,y,z)$ and valley, respectively, with the sublattice index implicit. 
    \begin{equation}\label{eqn:StagFluxMajoranaHopping-X}
      \begin{aligned}
        \delta H  &= -4 \delta \phi \sum_\vi \left[\chi_{0,1}^T(x_\vi) \rho^x + \left( -1 \right)^{i_y} \chi_{0,2}^T(x_\vi)\right] \left( -1 \right)^{i_y}\rho^x  \left[ \rho^x \chi_{z,1}(x_\vi) + \left( -1 \right)^{i_y} \chi_{z,2}(x_\vi) \right]
        \\
        &+2 \delta \phi \sum_\vi \left[\chi_{0,1}^T(x_\vi) \rho^x + \left( -1 \right)^{i_y} \chi_{0,2}^T(x_\vi)\right] \left( -1 \right)^{i_y}(\rho^x - i\rho^y)  \left[ \rho^x \chi_{z,1}(x_{\vi + \hat{\vx}}) + \left( -1 \right)^{i_y} \chi_{z,2}(x_{\vi+\hat{\vx}}) \right]
        \\
        &+2\delta \phi \sum_\vi \left[\chi_{0,1}^T(x_\vi) \rho^x + \left( -1 \right)^{i_y} \chi_{0,2}^T(x_\vi)\right] \left( -1 \right)^{i_y}(\rho^x + i \rho^y)  \left[ \rho^x \chi_{z,1}(x_{\vi -\hat{\vx}}) + \left( -1 \right)^{i_y} \chi_{z,2}(x_{\vi-\hat{\vx}}) \right]
        \\
        &\approx 2 \delta \phi \int\dd[2]{x} \left[ \chi_{0,1}^T \rho^z \partial_x \chi_{z,2} - \chi_{0,2}^T \rho^z \partial_x \chi_{z,1} \right]
        \\
        &\Rightarrow \delta \mathcal{L} = - 2 i  \delta \phi \Tr\left(\sigma^z \bar{X} \mu^y \gamma^y \partial_x X \right)
      \end{aligned}
    \end{equation}
    In our final term, we have reintroduced the $x \leftrightarrow y$ hoppings.
    For the hoppings in the $y$-direction, 
    \begin{equation}
    \label{eqn:StagFluxMajoranaHopping-Y}
      \begin{aligned}
        \delta H &=2 \delta \phi \sum_\vi \left[\chi_{0,1}^T(x_\vi) \rho^x + \left( -1 \right)^{\vi_y} \chi_{0,2}^T(x_\vi)\right] \left( -1 \right)^{i_y} 
         \left[ \rho^x \chi_{z,1}(x_{\vi + \hat{\vy}}) - \left( -1 \right)^{i_y} \chi_{z,2}(x_{\vi+\hat{\vy}}) \right]
        \\
        &-2 \delta \phi \sum_\vi \left[\chi_{0,1}^T(x_\vi) \rho^x + \left( -1 \right)^{i_y} \chi_{0,2}^T(x_\vi)\right] \left( -1 \right)^{i_y}  \left[ \rho^x \chi_{z,1}(x_{\vi - \hat{\vy}}) - \left( -1 \right)^{i_y} \chi_{z,2}(x_{\vi-\hat{\vy}}) \right]
        \\
        &\approx -2 \delta \phi \int \dd[2]{x} \left[ \chi^T_{0, 1} \rho^x \partial_y \chi_{z, 2} - \chi^T_{0,2} \rho^x \partial_y \chi_{z, 1} \right]
        \\
        &\Rightarrow \delta \mathcal{L} =-2 i \delta \phi \Tr\left(\sigma^z \bar{X} \mu^y \gamma^x \partial_y X \right)
      \end{aligned}
    \end{equation}
Note that here and in Eq.~\eqref{eqn:StagFluxMajoranaHopping-X} the Pauli matrix $\sigma^z$ is acted on by the SU(2) gauge symmetry of the $\pi$-flux phase.
Gauge invariance requires there exist nearly identical continuum model bilinears containing instead $\sigma^x$ and $\sigma^y$ Pauli matrices. 
It is therefore useful to express the perturbation in a gauge independent fashion using an
adjoint Higgs field $\Phi_3^a$, where $a=x,y,z$ is a SU(2) gauge index: 
    \begin{equation}
      \begin{aligned}
     \delta \mathcal{L} =   \Phi_3^a \Tr \left[ \sigma^a \bar{X} \mu^y \left( \gamma^y i \partial_x + \gamma^x i \partial_y \right) X\right]\,.
      \end{aligned}
    \end{equation}
(Our choice of subscript ``3" will be clear shortly.)
This Higgs field mediates the onset of the staggered flux phase, and in this present gauge we have the identification
\beq
\Phi_3^z \sim \delta \phi\,.
\eeq
Condensing the Higgs field with $\langle \Phi^a_3 \rangle$ leads to a transition from the $\SU(2)$ $\pi$-flux state to the U(1) staggered flux state. 
For concreteness, we continue to work in the gauge where $\Phi^a_3$ condenses in the $z$ direction, as implied by Eqs.~\eqref{eqn:StagFluxMajoranaHopping-X} and~\eqref{eqn:StagFluxMajoranaHopping-Y}.

\subsection{From $\pi$-flux to Z2A$zz13$}
\label{sec:pitoZ2}

We now evaluate the effects of a non-zero $\gamma_{1,2}$ in the $\pi$-flux phase, using the Majorana gauge as given in Eq.~\eqref{su2ansatz}. We first consider turning on the perturbation
\begin{equation}\label{eqn:NNNhopping}
  \begin{aligned}
    \bar{u}_{\vi, \vi+\hat{\vx} + \hat{\vy}} = \bar{u}_{\vi, \vi-\hat{\vx}+\hat{\vy}} = \gamma_1 \tau^y - \gamma_2 \tau^x \,,\quad i_x + i_y \text{ even}\,.
  \end{aligned}
\end{equation}
Recall that in the Majorana basis, terms proportional to $\tau^x$ ($\tau^y$) correspond to hoppings between the $0\leftrightarrow x (y)$ and $z\leftrightarrow y(x)$ Majorana fermions. Focusing on the $\gamma_1$ term, we expand in low-energy modes
\begin{equation}
  \begin{aligned}
    \delta H &= \gamma_1 \sum_\vi \left[ \chi_{0,1}^T(x_\vi) \rho^x + \left( -1 \right)^{i_y} \chi^T_{0,2}(x_\vi) \right] \left[ \rho^x + \left( -1 \right)^{i_y} \rho^y \right]\left[ \rho^x \chi_{y,1}(x_{\vi+\hat{x}+\hat{y}}) - \left( -1 \right)^{i_y} \chi_{y,2}\left( x_{\vi+\hat{x}+\hat{y}} \right) \right]
    \\
    &\approx \gamma_1 \sum_\vi \chi_{0,1}^T(x_\vi) \rho^x \chi_{y,1}(x_\vi) - \chi_{0,2}^T(x_\vi) \rho^x \chi_{y,2}(x_\vi) + \chi_{0,1}^T(x_\vi) \rho^z \chi_{y,2}(x_\vi) + \chi_{0,2}^T(x_\vi) \rho^z \chi_{y,1}(x_\vi)
    \\
    &\Rightarrow \delta \mathcal{L} = \gamma_1 \Tr\left[ \sigma^y \bar{X} \left( \mu^z \gamma^x +\mu^x \gamma^y \right)X \right]
  \end{aligned} \label{deltaH1}
\end{equation}
The perturbation is identical for the $\gamma_2$ term, but with $\sigma^y \Rightarrow - \sigma^x$.

As in the previous section, the addition of the hopping parameters of Eq.~\eqref{eqn:NNNhopping} can be formulated in a gauge-invariant fashion by coupling the bilinear above to an adjoint Higgs field $\Phi^a_{\bar{1}}$, $a=x,y,z$ (the bar on the ``1" will be apparent below). In particular, when a term $\Phi_{\bar{1}}^a \Tr\left[ \sigma^a \bar{X} \left( \mu^z \gamma^x +\mu^x \gamma^y \right) X \right]$ is added to the Lagrangian, we reproduce the continuum version of Eq.~\eqref{eqn:NNNhopping} we just derived when $\Phi_{\bar{1}}^a$ condenses as
\beq
\langle\Phi_{\bar{1}}^x\rangle  \sim \gamma_2
\quad , \quad \langle\Phi_{\bar{1}}^y\rangle \sim \gamma_1\,.
\eeq

We perform the same analysis for the second term proportional to $\gamma_{1,2}$:
\begin{equation}\label{eqn:NNNhopping-2}
  \begin{aligned}
    u_{\vi, \vi+\hat{\vx} + \hat{\vy}} = u_{\vi, \vi-\hat{\vx}+\hat{\vy}} = \gamma_1 \tau^x + \gamma_2 \tau^y \,,\quad i_x + i_y \text{ odd}\,.
  \end{aligned}
\end{equation}
The continuum derivation of this is essentially identical to as before, yielding
\begin{equation}
  \begin{aligned}
    \delta \mathcal{L} = \Tr\left[ \left(\gamma_1 \sigma^x + \gamma_2 \sigma^y \right) \bar{X} \left( \mu^z \gamma^x -\mu^x \gamma^y \right)X \right]\,,
  \end{aligned}
\end{equation}
prompting use to introduce $ \Phi_{\bar{2}}^a \Tr\left[\sigma^a  \bar{X} \left( \mu^z \gamma^x -\mu^x \gamma^y \right)X \right]$.
The continuum version of Eq.~\eqref{eqn:NNNhopping} is obtained through the condensation $\Phi_{\bar{2}}^a$ such that $\langle \Phi_{\bar{2}}^a \rangle = \gamma_1 \delta_{ax} + \gamma_2 \delta_{ay}$. 

\subsection{Majorana-Higgs Lagrangian}
\label{sec:MajoranaHiggs}

We now combine the results of Sections~\ref{sec:pifluxso5}, \ref{sec:pitostag}, and \ref{sec:pitoZ2} to obtain the low energy Lagrangian for the Majorana field $X$, and 3 real, adjoint Higgs scalars, which we now identify as $\Phi_{1}^a$, $\Phi_{2}^a$, $\Phi_{3}^a$ ($\Phi_{1,2}^a$ are rotations of $\Phi_{\bar{1},\bar{2}}^a$ in the $1,2$ plane). We do not explicitly write out the coupling to the SU(2) gauge field in this subsection, which can be included by the usual requirements of minimal coupling.

The Lagrangian is
\begin{equation}
  \begin{aligned}
    \mathcal{L} &= i \Tr \left( \bar{X} \gamma^\mu \partial_\mu X \right) + \Phi_1^a \Tr \left( \sigma^a \bar{X} \mu^z \gamma^x X \right)+ \Phi_2^a \Tr \left( \sigma^a \bar{X} \mu^x \gamma^y X \right)  \\ 
    &~~~~~~~~~~~~+ \Phi_3^a \Tr \left( \sigma^a \bar{X} \mu^y \left( \gamma^y i \partial_x + \gamma^x i \partial_y \right) X\right)\, + V (\Phi) .
  \end{aligned}
  \label{LMH}
\end{equation}
The staggered flux state is obtained when $\langle \Phi_3 \rangle \propto (0, 0, \delta \phi)$. The Z2A$zz13$ state follows from $\langle \Phi_1 \rangle \propto (\gamma_1 - \gamma_2, \gamma_1 + \gamma_2, 0)$ and $\langle \Phi_2 \rangle \propto (-\gamma_1 - \gamma_2, \gamma_1 - \gamma_2, 0)$. 

The Higgs potential $V(\Phi)$ arises from integrating out the high energy spinon degrees of freedom. We deduce its form by carefully considering the symmetry properties of the theory, which are described in some detail in Appendix~\ref{app:psg}.
Here, we note that the theory should respect time reversal and the lattice symmetries, 
\begin{align}
T_x&: (i_x,i_y)\mapsto (i_x+1,i_y),
&
T_y&: (i_x,i_y)\mapsto (i_x,i_y+1),
\nonumber \\
P_x&: (i_x,i_y) \mapsto (-i_x,i_y),
&
P_y&: (i_x,i_y)\mapsto (i_x,-i_y),
\nonumber \\
R_{\pi/2}&: (i_x,i_y)\mapsto (-i_y,i_x)\,,
\end{align}
and we summarize the transformations of the Higgs fieds here: 
\begin{center}
  \begin{tabular}{c||c|c|c|c|c|c}
    & $T_x$ & $T_y$ & $P_x$ & $P_y$ & $\mathcal{T}$ & $R_{\pi/2}$
    \\
    \hline\hline
    $\Phi_1^a$ & $-$ & $+$ & $-$ & $-$ & $-$ & $-\Phi_2^a$
    \\
    $\Phi_2^a$ & $+$ & $-$ & $-$ & $-$ & $-$ & $-\Phi_1^a$
    \\
    $\Phi_3^a$ &$-$ & $-$& $+$ & $+$ & $+$ & $-$
    \\
  \end{tabular}\,.
\end{center}
From this, we can deduce that the following gauge-invariant terms are allowed to quartic order in the Higgs potential
\bea
V(\Phi) &=& s \left( \Phi_1^a \Phi_1^a + \Phi_2^a \Phi_2^a \right) + \widetilde{s} \, \Phi_3^a \Phi_3^a + w \, \epsilon_{abc} \, \Phi_{1}^a \Phi_{2}^b \Phi_{3}^c \nonumber \\
&+& u \left( \Phi_1^a \Phi_1^a + \Phi_2^a \Phi_2^a \right)^2 + \widetilde{u} \left( \Phi_3^a \Phi_3^a \right)^2 + v_1  \left( \Phi_1^a \Phi_2^a \right)^2  +  v_2 \left( \Phi_1^a \Phi_1^a \right)\left( \Phi_2^b \Phi_2^b \right) \nonumber \\
 &+& v_3 \left[ \left( \Phi_1^a \Phi_3^a \right)^2  + \left( \Phi_2^a \Phi_3^a \right)^2 \right]  +  v_4 \left( \Phi_1^a \Phi_1^a + \Phi_2^a \Phi_2^a \right)\left( \Phi_3^b \Phi_3^b \right)\,.\label{lambda4}
\eea
where $\epsilon_{abc}$ is the antisymmetric unit tensor.

An important feature of $V(\Phi)$ is the cubic term proportional to $w$. This term implies that if any two of the Higgs fields are condensed, then so must the third. It also shows that even if we were only considering the transition from the SU(2) $\pi$-flux phase to the gapless $\mathbb{Z}_2$ spin liquid by the condensation of $\Phi_{1,2}^a$, we would be forced to include $\Phi_3^a$ in our theory, and hence the additional possibility of a U(1) staggered flux phase. The symmetry transformations show that $\Phi_3^a$ is the unique adjoint Higgs field that can be made from the tensor product of the Higgs fields needed to describe the gapless $\mathbb{Z}_2$ spin liquid, $\Phi_1^a$ and $\Phi_2^a$: so the staggered flux phase is a natural partner of this gapless $\mathbb{Z}_2$ spin liquid and the $\pi$-flux phase.

We can perform a mean-field minimization of Eq.~(\ref{lambda4}), and typical results are shown in Fig.~\ref{fig:mfPhaseDiagram}. There are 3 phases as a function of the tuning parameters $s$ and $\widetilde{s}$, which correspond to exactly those obtained in the lattice mean-field theory described in Section~\ref{sec:gaplessz2}.
The presence of the $w$ term implies that there is a first order transition line near the point where the 3 phases meet \cite{SSPark02}, as shown in Fig.~\ref{fig:mfPhaseDiagram}. 
We summarize and re-express the lattice theory results in terms of the continuum model parameters below.

\subsubsection{SU(2) $\pi$ flux phase}

Here, there is no Higgs condensate $\langle \Phi_{1,2}^a \rangle = 0$, $\langle \Phi_{3}^a \rangle = 0$,
and the system lies in the red region on the top right of Fig.~\ref{fig:mfPhaseDiagram}: the SU(2) $\pi$-flux phase.
The continuum model possesses an SU(2) gauge symmetry, along with the corresponding gauge bosons. 
The theory is believed to confine to  the N\'eel or VBS phase---as discussed in Section~\ref{sec:intro} and \ref{sec:conc}, we view the N\'eel phase to be more likely. 

\subsubsection{U(1) Staggered flux phase} This state as $\langle \Phi_3^a \rangle$ non-zero, while $\langle \Phi_{1,2}^a \rangle = 0$, resulting in the U(1) staggered flux phase represented on the top left of Fig.~\ref{fig:mfPhaseDiagram}.
Making contact with the lattice ansatz, we have 
\beq
\langle \Phi_3^a \rangle \propto (0,0,\phi - \pi/4) \neq 0\,.
\eeq
Again, the theory has a continuous unbroken gauge degree of freedom, now with only a U(1) symmetry.
There is a single gauge boson, which we nevertheless assume triggers confinement.
As argued, the most likely fate of the theory is the VBS state, but we cannot preclude the N\'eel phase.

\subsubsection{$\mathbb{Z}_2$ spin liquid Z2A$zz13$} 

The $\mathbb{Z}_2$ spin liquid Z2A$zz13$ corresponds to a Higgs condensate satisfying $\langle \Phi_{1,2}^a \rangle \neq 0$; it is shown in the lower half of the phase diagram of Fig.~\ref{fig:mfPhaseDiagram}.
The symmetry transformations imply that $\Phi_1^a$ and $\Phi_2^a$ have the same mass, so only a single tuning parameter, $s$,  is required to make them condense from the SU(2) $\pi$-flux phase. From the symmetry transformations, we also see that the absence of a broken symmetry requires that the gauge-invariant bilinears obey 
\beq 
\langle \Phi_1^a \Phi_1^a \rangle = \langle \Phi_2^a \Phi_2^a \rangle > 0, \quad , \quad \langle \Phi_1^a \Phi_2^a \rangle = 0. 
\eeq
Such saddle points are obtained from the Higgs potential for a range of $v_1$ positive and $v_2$ negative.
Moreover,  such a saddle point is indeed present in the lattice ansatz of the previous section where
\beq
\langle \Phi_{1a} \rangle \propto (-\gamma_1 - \gamma_2, \gamma_1 - \gamma_2,0) \quad, \quad \langle \Phi_{2a} \rangle \propto (\gamma_1 - \gamma_2, \gamma_1 + \gamma_2,0)\,. \label{eq:Phi12}
\eeq
We note that this implies $\langle\boldsymbol{\Phi}_1\rangle\perp\langle\boldsymbol{\Phi}_2\rangle$ and $|\langle\boldsymbol{\Phi}_1\rangle|=|\langle\boldsymbol{\Phi}_2\rangle|$, where we use a vector shorthand for the indices $a=x,y,z$ of the Higgs fields. 
By minimizing the potential $V(\Phi)$ in
Eq.~(\ref{lambda4}), we see that this $\mathbb{Z}_2$ spin liquid also implies the condensation of the remaining Higgs field: 
\beq 
\langle \Phi_{3}^a\rangle \propto w\, \epsilon_{abc} \, \langle \Phi_{1}^b \rangle \langle \Phi_{2}^c \rangle \label{Phi321}
\eeq
It follows that $\widetilde{s}$ can change sign within this phase without any phase transition.

\subsection{Visons}
\label{sec:visons}

The $\mathbb{Z}_2$ spin liquid is obtained from the theory in Eq.~\eqref{LMH} + SU(2) gauge fields (which is Eq.~\eqref{eq:fullLocalLagrangian} below) by condensing $\Phi_{1,2}^a$. This spin liquid has gapless fermionic spinon excitations, whose low energy dispersion can also be determined from the continuum theory. However, as in all $\mathbb{Z}_2$ spin liquids, there must also be vison excitations, which are mutual semions with respect to the spinons.
In the theory in Eq.~\eqref{eq:fullLocalLagrangian}, the vison is a finite energy excitation associated with vortex-like saddle point in which the Higgs fields $\Phi_{1,2}^a$ undergo a topologically non-trivial SO(3) rotation, associated with $\pi_1$(SO(3))$=\mathbb{Z}_2$, around the core of the vortex: see Ref.~\cite{SSST19} for an explicit solution in a theory without the fermionic spinons. Given that the vison appears in a lattice model with a background spinon density of one spinon per site, we expect the vison transforms projectively under translational symmetries with $T_x T_y = - T_y T_x$, where $T_\alpha$ is translation by one lattice spacing in the $\alpha$ direction \cite{RJSS91,MVSS99,TSMPAF99,Huh11,SSROPP}. For the case of gapped spinons, this fact now has a modern interpretation in the theory of symmetry fractionalization in topological phases \cite{Zaletel:2014epa,Cheng:2016pjt,Bonderson16,Metlitski:2017fmd,Else:2019lft}. We expect that a similar result applies in the present gapless spinon case, but this has not been explicitly established. For the case of gapped spinons, the vison projective transformation can be derived from a parent U(1) gauge theory (which is Higgsed down to $\mathbb{Z}_2$) in which the monopoles carry Berry phases \cite{RJSS91,MVSS99,SSROPP}. Such monopole Berry phases are in-turn related to a SO(5) Wess-Zumino-Witten term in an effective theory the N\'eel and VBS order parameters \cite{Tanaka_2005,SenthilFisher06}. Notably, this SO(5) WZW term
is also linked to an anomaly of the Majorana theory in Eq.~\eqref{eq:fullLocalLagrangian} \cite{Wang17}. It would therefore be interesting to establish $T_x T_y = - T_y T_x$ for gapped visons in the presence of gapless spinons starting directly from Eq.~\eqref{eq:fullLocalLagrangian} and condensing the Higgs fields: we leave such an analysis for future work.

\section{Renormalized perturbation expansion for the critical SU(2) gauge theory}
\label{sec:rg}

This section will present an analysis of the transition obtained by tuning the Higgs `mass' $s$ in Eq.~\eqref{lambda4} across a quantum critical point at $s=s_c$, for $\widetilde{s}>0$ in Fig.~\ref{fig:mfPhaseDiagram}, between the SU(2) and $\mathbb{Z}_2$ spin liquids. We have $\langle \Phi_{1,2,3}^a \rangle = 0$ for $s>s_c$, yielding the $\pi$-flux spin liquid. For $s<s_c $, we have $\langle \Phi_{1,2}^a \rangle \neq 0$ yielding the $\mathbb{Z}_2$ spin liquid Z2A$zz13$. As we noted below Eq.~\eqref{eq:Phi12}, $\langle \Phi_3^a \rangle$ will also be non-zero once both $\langle \Phi_{1,2}^a \rangle$ are non-zero. However, as $\langle \Phi_3^a \rangle$ is quadratic in $\langle \Phi_{1,2}^a \rangle$ (see Eq.~\eqref{Phi321}), it is not a primary order parameter for the transition. So we can entirely neglect $\Phi_3^a$ in the analysis of the criticality in the present section.

It is also convenient to write the theory in terms of 2 flavors of complex Dirac fermions which also carry a fundamental SU(2) gauge charge, $\psi_{\alpha,v}$; Here $\alpha$ is the SU(2) gauge index, $v=1,2$ is the valley index, and the Dirac/sublattice index is suppressed. The global SU(2) spin symmetry is not manifest in this formalism, unlike in the earlier Majorana formalism. Since the Lagrangian in Eq.~\ref{LMH} does not contain terms that act on the physical SU(2) spin, our Lagrangian nevertheless has a simple form in terms of these Dirac fermions, although a more careful analysis will be required to calculate the behavior of the N\'eel order parameter, which does involve the physical SU(2) spin. Explicitly, the relationship between the Dirac and Majorana fermions is
\beq
\psi_{\alpha, v} = i \sigma^y_{\alpha, \beta} X_{1, v;\beta}\,. 
\eeq
Applying this change of variables to Eq.~\eqref{LMH}, and including the SU(2) gauge field $A^a_\mu$, we obtain the Lagrangian for $\psi$ and the $\Phi_{1,2}^a$ Higgs fields
\begin{equation}
  \begin{aligned}
    \mathcal{L} &= \mathcal{L}_{\psi} + \mathcal{L}_{\Phi} + \mathcal{L}_{\Phi \psi}
    \\
    \mathcal{L}_{\psi} &= i \sum_{v} \bar{\psi}_v \gamma^\mu \left( \partial_\mu - i A^a_\mu \sigma^a \right) \psi_v\,.
    \\
    \frac{\mathcal{L}_{\Phi}}{N_f} &= \frac{K}{2} \left[ (\partial_x \Phi_1^a - 2\epsilon_{abc} A_x^b \Phi_1^c)^2 + (\partial_y \Phi_2^a - 2\epsilon_{abc} A_y^b \Phi_2^c)^2 \right] + \frac{s}{2} \left( \Phi_1^a \Phi_1^a + \Phi_2^a \Phi_2^a \right) \\
    & ~~~~~~~+
    u \left( \Phi_1^a \Phi_1^a + \Phi_2^a \Phi_2^a \right)^2  + v_1  \left( \Phi_1^a \Phi_2^a \right)^2  +  v_2 \left( \Phi_1^a \Phi_1^a \right)\left( \Phi_2^b \Phi_2^b \right)
    \\
    \mathcal{L}_{\Phi \psi} &= \lambda\left( \Phi_1^a \, \bar{\psi} \mu^z \gamma^x \sigma^a \psi + \Phi_2^a \, \bar{\psi} \mu^x \gamma^y \sigma^a \psi \right)
    \label{eq:fullLocalLagrangian}
  \end{aligned}
\end{equation}
We will henceforth work in Euclidean signature, with $\left(\gamma^\mu\right)^2 = 1$ for all $\mu$. This Lagrangian includes an important new term not present in Eq.~\eqref{LMH}: a bare spatial gradient term for the Higgs field proportional to the coupling $K$ (we will define $N_f$ shortly). This coupling is allowed by symmetry, and will turn out to be `dangerously irrelevant'
{\it i.e.\/} under renormalization, $K$ flows to zero, but it cannot be set to zero at the outset because of some singular effects that we will describe below. In contrast, the quartic couplings $u$, $v_{1,2}$ are geniunely irrelevant at the critical point, and will not be considered further.

The theory $\mathcal{L}$ is invariant under SU(2) gauge, SU(2) spin rotation, time-reversal, and space group transformations, as it must be, because these are symmetries of the underlying Hamiltonian and its parton representation. However, the Yukawa coupling $\lambda$ breaks both the emergent Lorentz and SO(5) symmetries of the fermion kinetic term. As we will show below, $\lambda$ is not an irrelevant perturbation, and so the absence of these emergent symmetries will be apparent in the critical correlation functions.

We will analyze the critical properties of Eq.~(\ref{eq:fullLocalLagrangian}) by the $1/N_f$ expansion used in earlier treatments of Dirac fermions coupled to scalar fields by Yukawa couplings which break relativistic invariance \cite{Huh08}. 
For this purpose, we will endow the fermions with an additional flavor index (not shown explicitly) which ranges over $N_f$ values. Combined with the $v$ index, there are a total of $2 N_f$ flavors and 2 colors of 2-component Dirac fermions. The physical case of interest to us is $N_f=1$.

As in Ref.~\cite{Huh08}, we will compute the renormalization constants of the theory $\mathcal{L}$ in a $1/N_f$ expansion. The most important of these will be the renormalization of the Fermi velocity, which has been implicitly set to unity above: this is non-zero because of the lack of the Lorentz invariance in the Yukawa coupling. The renormalization of the Fermi velocity in turn defines a dynamic critical exponent $z$: we will compute $z$ to order $1/N_f$ and find it to be a universal number at this order. Next, we shall examine the renormalization of the field scales.  As in the Ref.~\cite{Huh08}, a convenient choice, as we explain below, is to renormalize the boson field scale $\Phi$ so that the Yukawa coupling $\lambda = 1$; we will assume $\lambda = 1$ below. As usual, the renormalization of the fermion field, $Z_\psi$, is determined from the fermion self energy, which then determines a fermion anomalous dimension $\eta_\psi$.
Here we will find an unusual phenomenon, which is one of our main results: the value of $\eta_\psi$ is not universal at order $1/N_f$, but has a logarithmic dependence upon the irrelevant coupling $K$. Finally, we will also compute the renormalization of the fermion bilinears associated with the N\'eel and VBS order parameters: these are not equal to each other because the SO(5) symmetry is explicitly broken. 

\subsection{Boson propagators}

\begin{figure}
    \centering
 \includegraphics[width=2.5in]{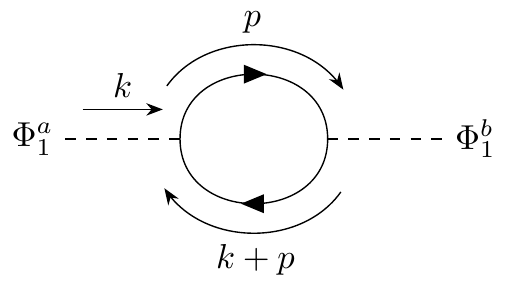}
  \includegraphics[width=2.5in]{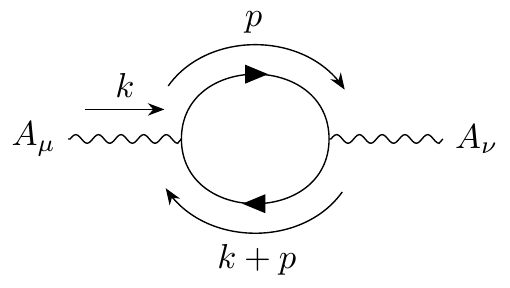}
    \caption{The leading order effective propagators for the Higgs (left) and gauge field (right) are generated by the one-loop contributions from $N$ fermions.}
    \label{fig:effectivePropagators}
\end{figure}

The first step in the large $N_f$ expansion is to integrate out the large number of fermions $\psi$, which allows us to determine the propagators of the bosons: the Higgs fields and the gauge fields. To leading order in $1/N_f$, we have to evaluate the diagrams in Fig.~\ref{fig:effectivePropagators}, and this leads to an effective quadratic action of the following form
\begin{equation}
  \begin{aligned}
    \frac{S_{b}}{N_f} &=  \int_k \frac{1}{2}(s + K k_x^2 + \Gamma_1(k)) \Phi_1^a (k) \Phi_1^a (-k) + \frac{1}{2}(s + K k_y^2 + \Gamma_2(k))\Phi_2^a (k) \Phi_2^a (-k)\nonumber \\
    &~~~~~~~~~~~~+  \frac{\Gamma_A(k)}{2} \left( \delta_{\mu\nu} - \frac{k_\mu k_\nu}{k^2} \right) A^a_\mu (k) A^a_\nu (-k) \,. \label{eq:bosonaction}
\end{aligned}
\end{equation}
We work in the Euclidean time signature, and $k$ is a 3-momentum.

We first calculate the one-loop corrections to the Higgs propagators. The correction to the $\Phi_1$ propagator is shown in the first diagram in Fig.~\ref{fig:effectivePropagators}, and is
\begin{equation}
  \begin{aligned}
    \Gamma_1(k) \delta_{ab} &=  \lambda^2 \Tr \int \frac{\dd[3]{p}}{(2\pi)^3}\left[ \mu^z \gamma^x \sigma^a \right] \left[ \frac{\slashed{p}}{p^2} \right]\left[ \mu^z \gamma^x \sigma^b \right]\left[ \frac{\slashed{p} + \slashed{k}}{(k+p)^2} \right] 
    \\
    &= - 8 \lambda^2 \delta_{ab} \int \frac{\dd[3]{p}}{(2\pi)^3}\frac{p_0 (p_0 + k_0) - p_x(p_x + k_x) + p_y (p_y + k_y)}{p^2 (k+p)^2}
\\
 &= \frac{ \lambda^2 \delta_{ab}}{4} \frac{(k_0^2 + k_y^2)}{\sqrt{k^2}}\cdot
  \end{aligned} \label{eq:gamma1}
\end{equation}
We have omitted a constant term, which will be tuned to zero at the critical point.

The correction to the $\Phi_2$ propagator is identical to the $\Phi_1$ correction, with $k_x \leftrightarrow k_y$.
\begin{equation}
  \begin{aligned}
    \Gamma_2(k) &= \frac{ \lambda^2}{4} \frac{(k_0^2 + k_x^2)}{\sqrt{k^2}}\cdot
    \label{eq:gamma2}
  \end{aligned}
\end{equation}

The reader should now notice some key features. As in Ref.~\cite{Huh08}, the overall scaling in momentum is $\Gamma_{1,2} \sim |k|$. So, this fermion-induced contribution to the $\Phi$ propagators is more important at low momenta than the $k^2$ terms which would be present in the bare theory. In general, the bare boson $k^2$ terms are irrelevant, and this is why we choose to set the field scale of $\Phi$ with the renormalization condition $\lambda=1$. However, unlike Ref.~\cite{Huh08}, we will see below in some detail that we cannot entirely ignore the bare $k^2$ term. The expression for $\Gamma_1$ ($\Gamma_2$) is not an increasing function of $k_x$ ($k_y$) when it is larger than the other momentum components, and this will lead to infrared singularities at first order in $1/N_f$. Specifically, the integral over the propagator $1/\Gamma_1$ ($1/\Gamma_2$) has an infrared divergence in the $k_0$,$k_y$ ($k_0$,$k_x$) plane. Consequently, we do need to include the {\it dangerously\/} irrelevant $K k_x^2$ ($K k_y^2$) term in the bare action for $\Phi_1^a$ ($\Phi_2^a$), as we have anticipated in Eqs.~\eqref{eq:fullLocalLagrangian} and \eqref{eq:bosonaction}. 

The $\order{N_f}$ propagator for the gauge field is obtained from 
\begin{equation}
  \begin{aligned}
    \Gamma_A (k) (k^2 \delta^{\mu} \delta^{\nu} - k^\mu k^\nu)  &= -    \Tr \int \frac{\dd[3]{p}}{(2\pi)^3} \gamma^\mu \left[ \frac{\slashed{p}}{p^2} \right] \gamma^\nu \left[ \frac{(\slashed{k} + \slashed{p})}{(k+p)^2} \right] 
\\
     &= \frac{1}{4  \sqrt{k^2}} (k^2 \delta^{\mu} \delta^{\nu} - k^\mu k^\nu) + \order{k^2}\,.
  \end{aligned}
\end{equation}
This is relativistically invariant, as expected.

\subsection{Fermion self-energy}
\begin{figure}
    \centering
    \includegraphics[width=2.5in]{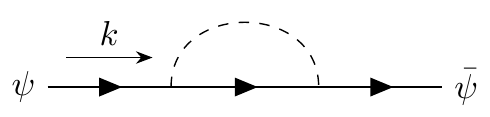}
    \includegraphics[width=2.5in]{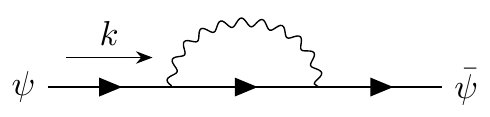}
    \caption{The two leading order contributions to the fermion self-energy, arising from Higgs (left) and gauge boson (right) couplings. To leading order in $1/N_f$, both the Higgs and gauge boson propagators are generated by the fermions.}
    \label{fig:selfEnergyDiagrams}
\end{figure}
We first calculate the one-loop corrections to the fermion self-energy, which will determine the anomalous dimension of the fermion operators as well as the dynamical critical exponent $z$. Although the anomalous dimension of the fermion is not a gauge-invariant observable, it will be needed to calculate the critical behavior of the gauge-invariant $\SO(5)$ order parameter. The three contributions to the fermion self-energy, as shown in Fig.~\ref{fig:selfEnergyDiagrams}, come from the two Higgs bosons and the gauge field, $\Sigma = \Sigma_1 + \Sigma_2 + \Sigma_A$.
\bea
    \Sigma_1(k) &=& \frac{3 }{N_f} \int \frac{\dd[3]{p}}{(2\pi)^3} \gamma^x \left[ \frac{\slashed{p} + \slashed{k}}{(p+k)^2} \right] \gamma^x  \frac{1}{\Gamma_1(p) + K p_x^2} \label{eq:Sigma1}
    \\
    \Sigma_2(k) &=& \frac{3 }{N_f} \int \frac{\dd[3]{p}}{(2\pi)^3} \gamma^y \left[ \frac{\slashed{p} + \slashed{k}}{(p+k)^2} \right] \gamma^y  \frac{1}{\Gamma_2(p) + K p_y^2} \label{eq:Sigma2}
    \\
    \Sigma_A(k) &=& \frac{3}{N_f} \int \frac{\dd[3]{p}}{(2\pi)^3} \gamma^\mu  \left[ \frac{\slashed{p} + \slashed{k}}{(p+k)^2} \right] \gamma^\nu \frac{\eta^{\mu\nu} - (1-\xi) \frac{p^\mu p^\nu}{p^2}}{\Gamma_A(p)p^2}\,.
    \label{eq:SigmaA}
\eea
We have introduced $\xi$ as a gauge-fixing parameter to obtain the gauge boson propagator.

Focusing on the Higgs corrections (Eqs.~\eqref{eq:Sigma1} and~\eqref{eq:Sigma2}), we analyze the behavior at small external momenta $k_i$. Note that the self-energy integrals are fully regulated by the presence of $K$ and a non-zero external momenta. Since $\Sigma_1$ ($\Sigma_2$) is invariant under $k_y \leftrightarrow k_0$ ($k_x \leftrightarrow k_0$), and the two transform into each other under a 90 degree spatial rotation, there are two distinct types of contributions for small external momenta. The first is proportional to $k_x \gamma^x$ for $\Sigma_1$, and $k_y \gamma^y$ for $\Sigma_2$. The second type includes all other possible choices of momenta, such as $k_0 \gamma^0$. 

As we shall justify below and in Appendix~\ref{app:rg}, we can focus on the regime $\abs{p_x} \gg \abs{p_0}, \abs{p_y}$ for graphs with a $\Phi_1$ propagator. In this limit, we can approximate the $\Phi_1$ propagator as
\begin{equation}
  \begin{aligned}
    \frac{4 \abs{p_x}}{p_0^2 + p_y^2 + 4 K \abs{p_x}^3}\,.
    \label{eq:higgsApprox}
  \end{aligned}
\end{equation}
At $K=0$, this propagator has an infrared divergence when integrated over the $p_{0,y}$ plane - so $K$ is needed an infrared regulator. 
With this, we extract the $\gamma^x$ correction to the self-energy from the $\Phi_1$ propagator by considering the $k_0=k_y=0$ limit:
\begin{equation}
    \gamma^x \Sigma_1(k_x) = \frac{12}{N_f} \int^\Lambda \frac{\dd[3]{p}}{(2\pi)^3} \frac{(p_x + k_x)}{(p_x + k_x)^2 + p_0^2 + p_y^2} 
    \frac{\abs{p_x}}{p_0^2 + p_y^2 + 4 K \abs{p_x}^3}\cdot
      \label{eq:selfEnergyLog2}
\end{equation}
We have indicated a cutoff $\Lambda$ to regulate the theory at large momenta, and this is needed in conformal gauge theories in 2+1 dimensions. However, with our inclusion of the irrelevant $K$ to control the infrared singularity, we find that the integrand vanishes faster at large momenta. It is not difficult to see that for $K \neq 0$ Eq.~(\ref{eq:selfEnergyLog2}) is finite as $\Lambda \rightarrow \infty$, and we will take this limit in the present section. The theory with a finite $\Lambda$ will be examined in Appendix~\ref{app:rg} in a renormalization group computation. 

We will now show that Eq.~(\ref{eq:selfEnergyLog2}) has a leading $k_x \ln^2 (k_x)$ contribution. One factor of $\ln (k_x)$ is the usual one: it follows from the fact that at $K=0$ the integrand divided by $k_x$ is a homogeneous function of momenta of dimension $-3$. The other comes from the infrared divergence regulated by $K$ noted below Eq.~(\ref{eq:higgsApprox}). 

Extracting the coefficient of the $k_x\ln^2(k_x)$ contribution requires a number of approximations. 
To understand the values of $p$ that dominate the integral in Eq.~(\ref{eq:selfEnergyLog2}), it is useful to perform the integral over $p_0$ and $p_y$:
\beq
  \gamma^x \Sigma_1(k_x)  =  \frac{12}{ N_f} \int \frac{\dd{p_x}}{8\pi^2} \frac{\abs{p_x} (p_x + k_x) \ln[(p_x + k_x)^2/(4K  \abs{p_x}^3)]}{(p_x + k_x)^2 - 4 K  \abs{p_x}^3}\,.
      \label{eq:selfEnergyLog2a}
\eeq
By examining the form of the integrals in Eqs.~\eqref{eq:selfEnergyLog2} and \eqref{eq:selfEnergyLog2a}, one can verify that the dominant term at small $k_x$ and $K$ is proportional to $k_x \ln^2( K k_x)$, and arises from the integration regime
\begin{equation}
  \begin{aligned}
    \left[ K \abs{p_x}^3 \right]^{1/2} \ll \left[ p_0^2 + p_y^2 \right]^{1/2} \ll \abs{p_x} \ll \frac{1}{ K}\,.
    \label{eq:logSquaredLimit}
  \end{aligned}
\end{equation}
The scale $K$ appears both as an ultraviolet cutoff and in defining the infrared bound.
For future calculations, this integration regime will prove to be the relevant one in isolating similar $\log^2$ contributions in other diagrams, although in principle one must still carry out an explicit calculation like in Eq.~\eqref{eq:selfEnergyLog2a} to verify that no other integration regimes give comparable contributions. We provide these calculations in Appendix~\ref{app:oneLoopCalcs} in addition to numerical evaluations of the one-loop integrals which confirm the validity of our approximations, and simply evaluate the one-loop integrals in the Eq.~\eqref{eq:logSquaredLimit} limit in the main text.

We can extract the coefficient of this $\log^2$ term by performing the integral in this regime,
\begin{equation}
  \begin{aligned}
   \gamma^x \Sigma_1 (k_x) & \approx  \frac{12}{N_f}   \int_{-1/K}^{1/K} \frac{\dd{p_x}}{2 \pi} \frac{|p_x|}{(p_x + k_x)}  \int_{(K|p_x|^3)^{1/2}}^{|p_x|} \frac{\dd{p_y} \dd{p_0}}{4 \pi^2} \frac{1}{p_0^2 + p_y^2}  
    \\
    & \approx  \frac{12}{N_f}   \int_{-1/K}^{1/K} \frac{\dd{p_x}}{2 \pi} \frac{|p_x|}{(p_x + k_x)} 
\frac{1}{4 \pi} \ln(1/(K |p_x|)) 
   \\
& \approx  - \frac{12}{N_f}  \frac{k_x}{8\pi^2} \left[ \ln(K k_x) \right]^2\,.
\end{aligned}
\end{equation}
Another discussion of the origin of the $k_x \ln^2 (k_x)$ is presented in Appendix~\ref{app:rg} using a renormalization group analysis. %

We now calculate the form of the second type of corrections using the limits in Eq.~\eqref{eq:logSquaredLimit}, evaluating the $\Phi_1$ contribution to the self-energy with external momentum $k_0$ for concreteness.
\begin{equation}
  \begin{aligned}
   \gamma^0 \Sigma_1(k_0) &\approx -\frac{12 k_0}{ N_f} \int^{1/K}_{-1/K} \frac{\dd{p_x}}{2\pi} \frac{1}{\abs{p_x}} \int_{(K\abs{p_x}^3)^{1/2}}^{\abs{p_x}} \frac{\dd{p_y} \dd{p_0}}{(2\pi)^2} \frac{1}{p_0^2 + p_y^2}
    \\
    &\approx -\frac{12 k_0}{N_f} \int^{1/K}_{-1/K} \frac{\dd{p_x}}{2\pi} \frac{1}{\abs{p_x}} \frac{1}{4\pi} \ln \left( 1/(K \abs{p_x}) \right) \approx -\frac{12}{N_f} \frac{k_0}{8\pi^2} \left[ \ln\left(K k_0 \right) \right]^2\,.
  \end{aligned}
\end{equation}
Combining the corrections from both Higgs propagators, we obtain the full expression for the self-energy for small external momenta at $\log^2$ order,
\begin{equation}
  \begin{aligned}
    \Sigma(k) \approx -\frac{3}{\pi^2 N_f}  \left[ k_0 \ln^2(K k_0) \gamma^0 + k_x \ln^2(K k_x) \gamma^x + k_y \ln^2(K k_y) \gamma^y \right]\,.
    \label{eq:selfEnergyCorrections}
  \end{aligned}
\end{equation}
In principle, the dependence on external momenta inside the logarithms could be more complicated for general $k$, i.e. $\Sigma(k) \gamma^0 \sim k_0 \ln^2(K f(k_0, k_x, k_y))$, but since we have verified that $f(k_0, 0, 0) = k_0$, then corrections to this are subleading.

These divergent corrections are absorbed into the renormalization of the fermion field, $\psi = \sqrt{Z_\psi} \psi_R$, with
\begin{equation}
    Z_\psi = 1 - \frac{3}{\pi^2 N_f} \ln^2(K \mu)\,, \label{eq:Zpsi}
\end{equation}
where we have renormalized the theory at some momentum scale $\mu$. This counterterm only cures the divergence at $\log^2$ order, since the renormalized self-energy at some other momentum scale $k$ will scale as
\begin{equation}
    \ln^2 (K \mu) - \ln^2(K k) = \ln(\mu/k) \ln(K^2 k \mu)\,.
\end{equation}
This, along with the RG analysis in Appendix~\ref{app:rg}, indicates that the subleading single-logarithm corrections will generically give non-universal behavior. However, these $\log^2$ corrections to the self-energy are Lorentz invariant, and do not affect the renormalization of the dynamical critical exponent, $z$. Therefore, the subleading single-logarithm correction to the velocity anistropy will lead to a universal correction to the dynamical critical exponent. To extract the subleading correction to $z$ using $K$ and the external momenta as a regulator, we start with the expression
\begin{equation}
  \begin{aligned}
    \pdv{\Sigma}{k_0} \gamma^0 - \pdv{\Sigma}{k_x} \gamma^x = - \frac{12}{N_f} \int \frac{\dd[3]{p}}{(2\pi)^3}\frac{2 (p_y+k_y)^2 }{(p+k)^4} \frac{\abs{p}}{4K p_y^2 \abs{p} + p_0^2 + p_y^2}\,.
    \label{eq:anisotropyOneLoop}
  \end{aligned}
\end{equation}
To leading order in $k$, we set $k=0$ inside the integrand and simply use it as an IR cutoff, which gives
\begin{equation}
  \begin{aligned}
    \pdv{\Sigma}{k_0} \gamma^0 - \pdv{\Sigma}{k_x} \gamma^x \approx  \frac{6}{ N_f \pi^2} \ln(K k)\,.
  \end{aligned}
\end{equation}
This result can be obtained analytically by approximating the integration region $k \leq \abs{p_x} \leq 1/K$, and can be verified by a numerical evaluation of Eq.~\ref{eq:anisotropyOneLoop}. This implies a renormalization of the Fermi velocity, $v_F = Z_{v} v_{F, R}$
\begin{equation}
    Z_{v} = 1 + \frac{6}{\pi^2 N_f} \ln\left(K \mu\right)
\end{equation}
The logarithmic derivative with respect to $1/K$
determines the renormalization of the dynamical critical exponent,
\begin{equation}
    z = 1 + \frac{6}{\pi^2 N_f}\,. \label{eq:zval}
\end{equation}
The one-loop calculation defined in Eq.~\eqref{eq:anisotropyOneLoop} is actually well-defined when $K=0$ and can be regulated via more standard approaches, such as dimensional regularization, as shown in Appendix \ref{app:dimreg}. The same value of $z$ is also obtained in a renormalization group computation in Appendix~\ref{app:rg}.

\subsection{\texorpdfstring{$\SO(5)$}{SO(5)} order parameter}
\begin{figure}
    \centering
  \includegraphics[width=1.5in]{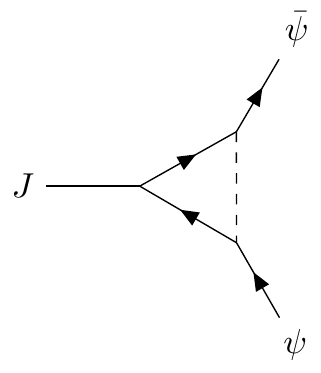}
  \includegraphics[width=1.5in]{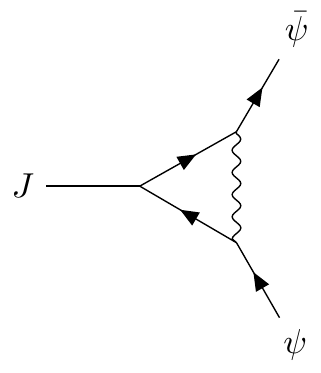}
  \includegraphics[width=2.3in]{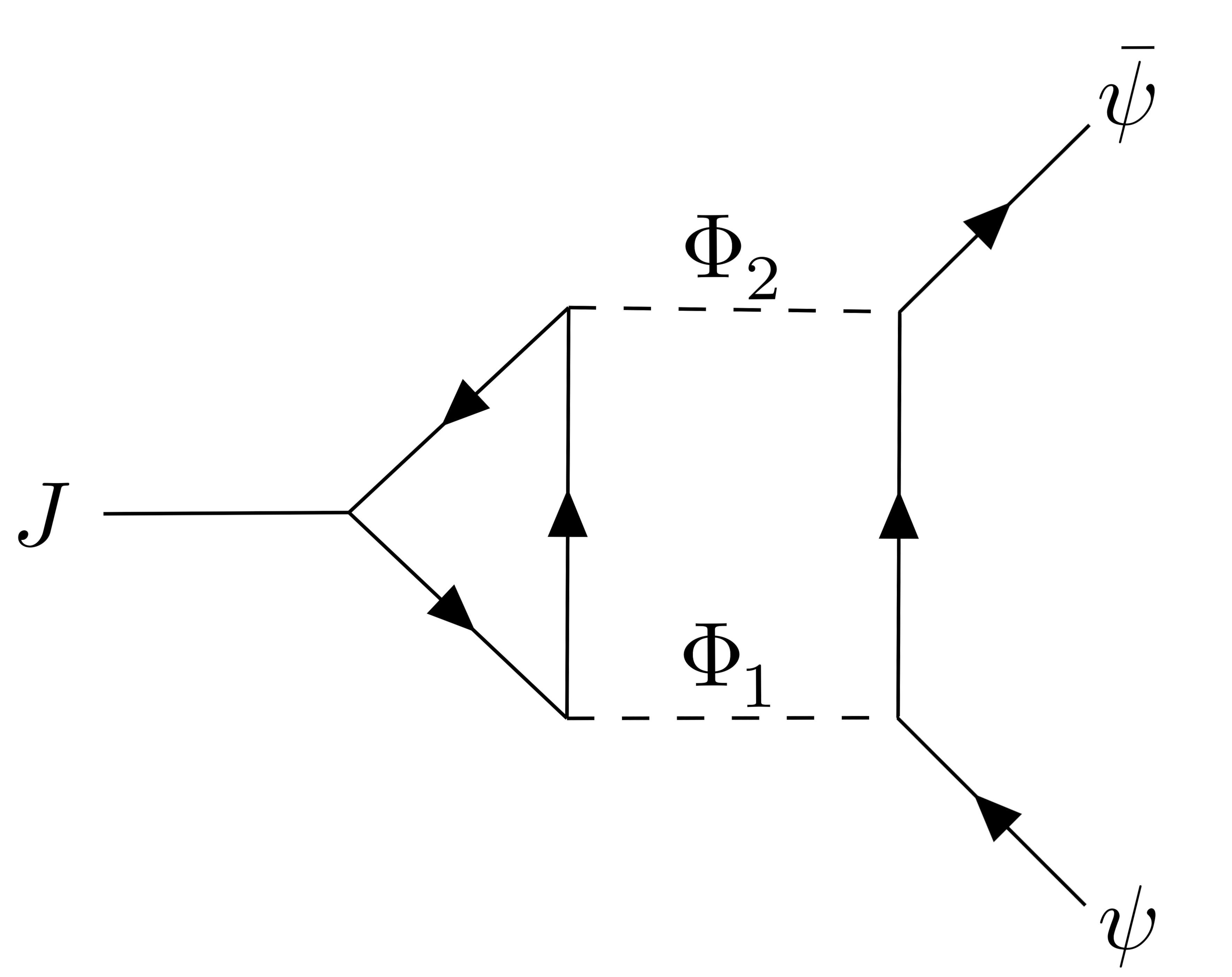}
    \caption{The $\mathcal{O}(N_f^{-1})$ vertex corrections which contribute to the renormalization of the $\SO(5)$ order parameter. The order parameter receives corrections at one-loop order from the Higgs fields (left) and the gauge boson (center), although only the former gives a $\log^2$ correction. An additional two-loop $\mathcal{O}(N_f^{-1})$ contribution (right) is possible - we show in Appendix~\ref{app:twoLoop} that it does not contain any $\log^2$ divergences.}
    \label{fig:so5diagrams}
\end{figure}

In the absence of the Higgs fields, our theory possesses an emergent $\SO(5)$ symmetry corresponding to rotations between N\'eel and VBS order parameters. This $\SO(5)$ symmetry is broken by the critical Higgs fields, and as a result, the scaling behavior of N\'eel and VBS order parameters will differ. In terms of Dirac fermions, the fermion bilinears corresponding to the two-component VBS order parameter - determined by the action of the square lattice symmetries on the bilinears - may be written as
\begin{equation}
  \begin{aligned}
    V^i = \bar{\psi} \Gamma^i \psi \,, \quad \Gamma^i = \{ \mu^x\,,\, \mu^z \}\,.
  \end{aligned}
\end{equation}
The three-component N\'eel order parameter has a less concise expression in terms of Dirac fermions - this is due to the fact that the Dirac fermion representation obfuscates the action of the physical $\SU(2)$ spin rotation symmetry. In terms of the Majorana field $X$, the order parameter is $\Tr\left(\bar{X} \mu^y \sigma^a X\right)$, $a = x\,,y\,,z$. In order to calculate corrections to the N\'eel order parameter, we focus on the $\sigma^z$ component, which happens to be simply expressible in terms of a Dirac fermion bilinear:
\begin{equation}
\begin{aligned}
N^z = \bar{\psi} \mu^y \psi\,. 
\end{aligned}
\end{equation}
Because the Higgs couplings preserve the physical $\SU(2)$ spin rotation symmetry, the other components must have the same corrections, and this has been confirmed by an explicit calculation in terms of the Majorana fermions. 
To compute the corrections to the scaling dimensions of these composite operators, we couple the fermion bilinear
$n^i =  \bar{\psi} \mu^i \psi$ to a source field $J_i$, and compute the $\order{N_f^{-1}}$ vertex corrections in Fig.~\ref{fig:so5diagrams}. 
Aside from the corrections coming from the renormalization of the fermion self-energy, the $\order{N_f^{-1}}$ corrections that we will be interested in come from one-loop corrections of the Higgs fields with external momenta $k_{1, 2}$
\begin{equation}
  \begin{aligned}
    &\frac{\mu^z \sigma^a \mu^i \sigma^a \mu^z}{N_f} \int \frac{\dd[3]{p}}{(2\pi)^3} \gamma^x \frac{\slashed{p} - \slashed{k_1}}{(p - k_1)^2} \frac{\slashed{p} - \slashed{k_2}}{(p - k_2)^2} \gamma^x \frac{1}{\Gamma_1(p) + K p_x^2}
    \\
    &+ \frac{\mu^x \sigma^a \mu^i \sigma^a \mu^x}{N_f} \int \frac{\dd[3]{p}}{(2\pi)^3} \gamma^y \frac{\slashed{p} - \slashed{k_1}}{(p - k_1)^2} \frac{\slashed{p} - \slashed{k_2}}{(p - k_2)^2} \gamma^y \frac{1}{\Gamma_2(p) + K p_y^2}\,,
  \end{aligned}
\end{equation}
where the first and second terms arise from interactions with $\Phi_1^a$ and $\Phi_2^a$, respectively. 
The gauge field correction does not break $\SO(5)$ symmetry and does not contribute to the renormalization at $\log^2$ order, so we will focus on the Higgs corrections. Additionally, there is a possible two-loop diagram shown in Fig.~\ref{fig:so5diagrams} that contributes at $\order{N_f^{-1}}$, but we show explicitly in Appendix~\ref{app:twoLoop} that these corrections also do not contribute to the renormalization at $\log^2$ order.
At zero external momenta, the $\log^2$ Higgs corrections to the VBS order parameter ($\mu^i = \mu^x\,, \mu^z$) drops out entirely, leaving only Higgs corrections coming from the fermion renormalization.

We focus on vertex corrections to the N\'eel order parameter ($\mu^i = \mu^y$). As is the case in the fermion self-energy, the spatial anisotropy in the Higgs propagators gives rise to $\log^2$ divergences in their corrections to the $\SO(5)$ vertex. We isolate $\log^2$ divergences in the Higgs correction to the $\SO(5)$ order parameter by including an external momenta $2k_x$ to the order parameter, which is distributed symmetrically between the two fermion fields. We calculate this for the $\Phi_1$ propagator---approximating the Higgs propagator as ${4\abs{p_x}}/({p_0^2 + p_y^2 + 4 K \abs{p_x}^3})$ as in the previous section and taking the limit in Eq.~\eqref{eq:logSquaredLimit}, the one-loop correction is
\begin{equation}
  \begin{aligned}
   & \frac{\mu^z \sigma^a \mu^y \sigma^a \mu^z}{N_f} \int \frac{\dd[3]{p}}{(2\pi)^3} \frac{p_x^2 - k_x^2 + p_y^2 + p_0^2}{\left[ (p_x + k_x)^2 + p_y^2 + p_0^2 \right] \left[ (p_x - k_x)^2 + p_y^2 + p_0^2 \right]} \frac{4\abs{p_x}}{p_0^2 + p_y^2+4 K \abs{p_x}^3}
   \\
   &\approx -\frac{12 \mu^y}{N_f} \int^{1/K}_{-1/K} \frac{\dd{p_x}}{2\pi} \frac{(p_x^2 - k_x^2) \abs{p_x}}{(p_x + k_x)^2 (p_x - k_x)^2} \int_{(K \abs{p_x}^3)^{1/2}}^{\abs{p_x}} \frac{\dd{p_y} \dd{p_0}}{4\pi^2} \frac{1}{p_0^2 + p_y^2}
   \\
   &\approx -\frac{12 \mu^y}{N_f} \int^{1/K}_{1/K} \frac{\dd{p_x}}{8\pi^2} \frac{(p_x^2 - k_x^2) \abs{p_x}}{(p_x + k_x)^2 (p_x - k_x)^2} \ln\left( 1/(K \abs{p_x}) \right) \approx  -\frac{3 \mu^y}{2 N_f\pi^2} \ln^2\left( K k_x \right)
   \label{eq:so5logSquaredIsolation}
  \end{aligned}
\end{equation}
The $\Phi_2$ propagator gives an identical correction.
Since the external momenta only play the role of an IR cutoff to leading order, we generalize this result to an arbitrary external momentum and obtain the composite operator renormalizations \cite{zinn2002quantum}
\begin{equation}
\begin{aligned}
    Z_{\text{VBS}} &= 1 
    \\
    Z_{\text{N{\'e}el}} &= 1  + \frac{3}{N_f \pi^2} \ln^2(K \mu) \,.
    \end{aligned}
    \label{eq:Zso5}
\end{equation}

We can state these results in terms of the perturbative corrections to the two-point correlator of the order parameters, $\langle \bar{\psi}\Gamma^i \psi (k)\bar{\psi}\Gamma^j \psi(-k)\rangle$, {\it i.e.\/} the corresponding susceptibilities $\chi_{\text{VBS}}$ and $\chi_{\text{N{\'e}el}}$; these combine the consequences of the composite operator renormalizations in Eq.~(\ref{eq:Zso5}), and $Z_\psi$ in Eq.~(\ref{eq:Zpsi}), to yield
\begin{equation}
\begin{aligned}
    \chi_{\text{VBS}} (k) & \sim -|k| \left(
    \frac{Z_\psi}{Z_{\text{VBS}}} \right)^2 =
     - |k|\left [ 1 - \frac{6}{N_f \pi^2} \ln^2 (K|k|) \right]
    \\
    \chi_{\text{N{\'e}el}} (k) & \sim -|k| \left(
    \frac{Z_\psi}{Z_{\text{N\'eel}}} \right)^2 = - |k| \left [ 1 - \frac{12}{N_f \pi^2} \ln^2 (K|k|) \right]\,.
    \end{aligned} \label{chik}
\end{equation}
After a Fourier transform to real space, these correlators are
\begin{equation}
\begin{aligned}
    \chi_{\text{VBS}} (r) & \sim  \frac{1}{|r|^4}\left [ 1 - \frac{6}{N_f \pi^2} \ln^2 (|r|/K) \right]
    \\
    \chi_{\text{N{\'e}el}} (r) & \sim  \frac{1}{|r|^4} \left [ 1 - \frac{12}{N_f \pi^2} \ln^2 (|r|/K) \right]\,.
    \end{aligned} \label{chir}
\end{equation}
The renormalization group analysis in Appendix~\ref{app:rg} shows how the above results may be renormalized to large $r$; we find
\begin{equation}
\begin{aligned}
    \chi_{\text{VBS}} (r) & \sim  \frac{1}{|r|^a}\exp \left( - \frac{6}{N_f \pi^2} \ln^2 (|r|/K) \right)
    \\
    \chi_{\text{N{\'e}el}} (r) & \sim  \frac{1}{|r|^b} \exp \left( - \frac{12}{N_f \pi^2} \ln^2 (|r|/K) \right)\,.
    \end{aligned} \label{chirg}
\end{equation}
where the exponents of the prefactors, $a$ and $b$, are {\it non-universal} numbers.

Leading logarithm-squared corrections have appeared earlier in a few other problems in quantum many-body theory. They appear in the theory of weakly disordered two-dimensional metals with Coulomb interactions \cite{AltshulerLee80,Finkelstein1984,LeeTVR}. More recently, $\log^2$ terms have also been found in computations of the density of states of clean bilayer graphene with Coulomb interactions \cite{Barlas09,Nandkishore10}. Renormalization group analyses of these cases \cite{Finkelstein1984,Nandkishore10} also yield an exponentiation similar to that in Eq.~(\ref{chirg}).

As an aside, we note that the one-loop vertex corrections to the bilinear $\bar{\psi} \psi$, whose symmetry properties identify it as the scalar spin chirality~\cite{Hermele05}, have the same magnitude and opposite sign as the N\'eel order parameter. Because of this, the $\log^2$ divergence is in fact cancelled by the fermion self-energy. As shown in Appendix~\ref{app:twoLoop}, the two-loop corrections coming from the Higgs fields vanish, meaning that correlations of the scalar spin chirality should have power law decay at $\order{N_f^{-1}}$. Since this power law decay is slower than the N\'eel and VBS correlations, this may indicate proximity to a chiral spin liquid. 
\section{Transition from U(1) staggered flux to gapless $\mathbb{Z}_2$ spin liquid}
\label{sec:sf}

This section discusses the critical U(1) gauge theory for the transition between the U(1) staggered flux spin liquid and the gapless $\mathbb{Z}_2$ spin liquid Z2$Azz13$ in Fig.~\ref{fig:mfPhaseDiagram}. A similar theory has been considered earlier \cite{SenthilLee05} for the N\'eel-$\mathbb{Z}_2$ spin liquid transition.

Both phases have the Higgs field $\langle \Phi_3^a \rangle \neq 0$. So let us fix $\Phi_3^a = \delta_{az} \Phi $, with $\Phi$ a non-zero constant, which will turn into a coupling constant in the low energy theory below.
In this situation, the SU(2) gauge symmetry is broken down to U(1), and we need only consider a U(1) gauge theory with the U(1) gauge field $A_\mu \equiv A^z_\mu$. Also important is the consequence of the $w$ term in the Higgs potential Eq.~(\ref{lambda4}):
\beq
V(\Phi) = \ldots + w \, \Phi\, (\Phi_1^x \Phi_2^y - \Phi_1^y \Phi_2^x )+ \ldots\,.
\eeq
Choosing a gauge with $w\, \Phi <0$, and diagonalizing the quadratic form of the Higgs potential for $\Phi_{1,2}^{x,y}$, we deduce that we need only focus on a single low energy complex Higgs field near the critical point
\beq
\mathcal{H} = \frac{1}{2} \left( \Phi_1^x + \Phi_2^y + i (\Phi_1^y - \Phi_2^x) \right) \,.
\eeq
It can now be checked that $\mathcal{H}$ transforms as a charge 2 Higgs field under the unbroken U(1) gauge symmetry. Other linear combinations of $\Phi_{1,2}^{x,y}$ can be ignored for the critical theory.

We can now obtain the critical theory for the fermions $\psi$, the complex Higgs field $\mathcal{H}$, and the U(1) gauge field $A_\mu$ from Eq.~(\ref{LMH}):
\begin{equation}
  \begin{aligned}
    \mathcal{L}_{\rm sf} &= \mathcal{L}_{\psi} + \mathcal{L}_{\mathcal{H}} + \mathcal{L}_{\mathcal{H} \psi}
    \\
    \mathcal{L}_{\psi} &= i \sum_{v} \bar{\psi}_v \gamma^\mu D_\mu \psi_v +  \Phi \, \bar{\psi} \mu^y \sigma^z \left(\gamma^y D_x + \gamma^x D_y \right) \psi \,.
    \\
    \frac{\mathcal{L}_{\mathcal{H}}}{N_f} &= s |\mathcal{H}|^2 + u |\mathcal{H}|^4
    \\
    \mathcal{L}_{\mathcal{H} \psi} &= \lambda \left( \mathcal{H} \bar{\psi} \left(\mu^z \gamma^x + i \mu^x \gamma^y \right) \sigma^-  \psi + \mathcal{H}^\ast \bar{\psi} \left(\mu^z \gamma^x - i \mu^x \gamma^y \right) \sigma^+  \psi \right)\,.
    \label{eq:sfLagrangian}
  \end{aligned}
\end{equation}
We define the covariant derivative $D_\mu = \partial_\mu - i A_\mu \sigma^z$ and operators $\sigma^{\pm} = ({\sigma^x \pm i \sigma^y})/{2}$.
Note that $\Phi$ is a marginal coupling constant here, not a fluctuating field. A crucial feature of $\mathcal{L}_{\rm sf}$ is that it does not contain the $K$ gradient terms: these terms are now truly irrelevant. This can be seen in the large $N_f$ expansion: upon integrating the fermions, we obtain, in place of Eq.~(\ref{eq:bosonaction}),
\begin{equation}
  \begin{aligned}
    \frac{S_{b}}{N_f} &=  \int_k \left(s + \frac{\Gamma_1(k)+ \Gamma_2 (k)}{2} \right)|\mathcal{H}(k)|^2 +  \frac{\Gamma_A(k)}{2} \left( \delta_{\mu\nu} - \frac{k_\mu k_\nu}{k^2} \right) A_\mu(k) A_\nu(-k) \,. \label{eq:bosonactionsf}
\end{aligned}
\end{equation}
where $\Gamma_{1,2} (k)$ are specified in Eqs.~(\ref{eq:gamma1}, \ref{eq:gamma2}) for $\Phi=0$. In general, the sum $\Gamma_1 (k)+\Gamma_2 (k)$ has the rotational symmetry of the square lattice, and its inverse does not contain the infrared singularities we encountered earlier. Consequently, there is no logarithmic violation of scaling by a dangerously irrelevant $K$ here, and the $1/N_f$ expansion of $\mathcal{L}_{\rm sf}$ should proceed along more conventional lines. 

The $1/N_f$ expansion of the theory $\mathcal{L}_{\psi}$ was presented in Refs.~\cite{RantnerWen02,Hermele05}: they found a stable Lorentz invariant fixed point with $\Phi=0$ at the fixed point. In our case, for $\mathcal{L}_{\rm sf}$ we 
expect a critical theory with dynamic scaling with an exponent $z \neq 1$, SO(5) symmetry broken by $\mathcal{L}_{\mathcal{H}\psi}$, and a spatial anisotropy in the fermion velocities at the Dirac nodes determined by the fixed point value of $\Phi$. Note that even for $\Phi=0$ we do not expect Lorentz invariance with $z=1$, because the relevant Yukawa couplings in $\mathcal{L}_{\mathcal{H} \psi}$ are not Lorentz invariant, and consequently $\Gamma_1 (k)+ \Gamma_2 (k)$ is not Lorentz invariant.

\section{Conclusions}
\label{sec:conc}

Building upon the results of recent numerical studies \cite{Sandvik18,Becca20,Imada20,Gu20}, we have proposed resolutions of long-standing controversies connected to theories of the cuprates: the phases of the frustrated square lattice spin $S=1/2$ antiferromagnets, and the nature of deconfined criticality in such models. Deconfined criticality expresses the low energy physics in terms of fractionalized degrees of freedom and emergent gauge fields, which can enter various confining states with possible broken symmetries on either or both sides of the critical point. Although there are several well-established examples, the transition between N\'eel and VBS states in square lattice antiferromagnets \cite{NRSS89,NRSS90,RS91,SR91} has been of particular interest. One formulation of this deconfined critical point is 
a version of QCD$_3$, quantum chromodynamics in 2+1 dimensions: a SU(2) gauge theory with 2 flavors of 2-component massless Dirac fermions, each carrying a fundamental color charge. This theory is dual to a SO(5) non-linear sigma model with a Wess-Zumino-Witten term \cite{Tanaka_2005,SenthilFisher06,Wang17}. There is now significant numerical evidence that such a conformal field theory (CFT) does not exist, although there is likely a 
nearby `complex' CFT \cite{Wang19,Nahum19,Assaad21,He20,Gorbenko:2018ncu,Gorbenko:2018dtm,Ma:2018euv}.
This leaves open the fate of a physical model with a Hermitian Hamiltonian, such as the $J_1$-$J_2$ antiferromagnet on the square lattice, between the N\'eel and VBS states. Here we have presented a theory in which the putative QCD$_3$ CFT is resolved into an intermediate stable gapless phase with $\mathbb{Z}_2$ topological order and gapless Dirac fermions \cite{TSMPAF99,WenPSG,Kitaev2006}. The intermediate $\mathbb{Z}_2$ spin liquid is flanked by two proposed deconfined critical points, neither of which is a CFT, or even invariant under Lorentz tranformations. 
The absence of Lorentz symmetry permits several novel phenomena, including the appearance of dangerously irrelevant couplings and logarithm-squared renormalizations, which can be tested in numerical studies. All of these phases and critical points are described by extending QCD$_3$ with 3 real adjoint Higgs fields. The couplings of these Higgs fields are tightly constrained by the transformations of QCD$_3$ under the symmetries of the underlying square lattice antiferromagnet, and an analysis of these symmetries occupy a significant portion of this paper.

Our main results can be summarized in the context of the mean-field phase diagram in Fig.~\ref{fig:mfPhaseDiagram} obtained from the SU(2) gauge theory with 3 adjoint Higgs field $\Phi_{1,2,3}^a$ in Eq.~(\ref{LMH}). This mean field theory yields 3 spin liquids, with deconfined SU(2), U(1), and $\mathbb{Z}_2$ gauge fields. We assume that the spin liquids with continuous gauge symmetries confine, except at possible deconfined critical transitions to the $\mathbb{Z}_2$ spin liquid. This phase diagram maps onto the $J_1$-$J_2$ model along the trajectory of the dotted blue line, and our proposed deconfined critical theories are at the boundaries between the mean field SU(2) and $\mathbb{Z}_2$ spin liquids, and the U(1) and $\mathbb{Z}_2$ spin liquids.

The numerical evidence for the confinement of the SU(2) $\pi$-flux spin liquid was reviewed in Section~\ref{sec:intro}. This confining state should have either N\'eel or VBS order \cite{Wang17}, and Ref.~\cite{Thomson17} argued by comparing to bosonic spinon theories that it should be the N\'eel state. 
The structure of the critical theory from such a confining state to the gapless $\mathbb{Z}_2$ spin liquid was presented in Section~\ref{sec:rg}, and we found some unusual log$^2$ corrections to both the N\'eel and VBS critical correlators.
From the geometry of the mean field phase diagrams in Fig.~\ref{fig:mfPhaseDiagram}, and the numerical studies on the square lattice antiferromagnet noted in Fig.~\ref{fig:becca}, it is then natural to propose that the U(1) staggered flux spin liquid confines to the VBS state. The critical U(1) gauge theory for the boundary between the U(1) and $\mathbb{Z}_2$ spin liquid was presented in Section~\ref{sec:sf}, and this has no log$^2$ terms. We also note that the log$^2$ correlators in Eqs.~(\ref{chir}) and (\ref{chirg}) show a faster decay of the N\'eel order than the VBS order, which might be evidence that the SU(2) critical theory is proximate to the VBS state rather than the N\'eel state, which would reverse the direction of the arrow in Fig.~\ref{fig:mfPhaseDiagram}.


Irrespective of the assignment of the N\'eel or VBS confining states to the SU(2) or U(1) spin liquids in Fig.~\ref{fig:mfPhaseDiagram}, we expect any direct phase boundary between the N\'eel and VBS states to be a first order transition. This follows from the numerical studies \cite{Wang19,Nahum19,He20} noted in Section~\ref{sec:intro}.

Our critical SU(2) gauge theory for the $S=1/2$ square lattice antiferromagnet has massless 2-component Dirac fermions with 2 flavors and 2 colors, and real critical Higgs fields with 2 flavors and 3 colors, and is shown in Eq.~\eqref{eq:fullLocalLagrangian}. This derives from a theory for the $\pi$-flux to gapless $\mathbb{Z}_2$ spin liquid transition proposed by Ran and Wen \cite{RanWen06,YingRanThesis}, and includes an additional `dangerously irrelevant' coupling $K$, which is the coefficient of a spatial gradient term in the Higgs fields. 
We analyzed this theory along the lines of the $1/N_f$ expansion of Ref.~\cite{Huh08} (the case of interest to us here is $N_f=1$). We found that the theory with $K=0$ has infrared divergencies that arise from the highly anisotropic spatial structure of the Higgs correlations, which is in turn a consequence of the non-Lorentz invariant Yukawa couplings between the Higgs fields and the fermions. So even though the coupling $K$ is formally irrelevant, it must be included to understand the long-distance and long-time behavior of the theory {\it i.e.\/} the coupling $K$ is dangerously irrelevant. We found that the coupling $K$ leads to leading logarithm-squared corrections to various correlators, such as those in Eqs.~(\ref{chik}) and (\ref{chir}) for the correlations of the N\'eel and VBS order parameters; Appendix~\ref{app:rg} showed how these corrections are exponentiated in a renormalization group analysis, lead to Eq.~(\ref{chirg}).
We also note that the logarithm-squared term was absent in the contributions to the dynamic critical exponent, $z$, and we computed a non-Lorentz-invariant value for $z$ in Eq.~\eqref{eq:zval}.

The critical U(1) gauge theory for the $S=1/2$ square lattice antiferromagnet was discussed in Section~\ref{sec:sf}. It has massless 2-component Dirac fermions with 4 flavors and $\pm 1$ U(1) gauge charges, and a single complex critical Higgs fields with $\pm 2$ U(1) gauge charge. We found that $K$ was not dangerously irrelevant in this theory.
The critical theory is not Lorentz invariant, and so has dynamic critical exponent $z \neq 1$. The critical theory also does not have the SO(5) symmetry between the N\'eel and VBS order parameters.
A full analysis of this theory requires a study of the role of anisotropies in the Dirac fermion velocities (associated with the coupling $\Phi$ in Eq.~\eqref{eq:sfLagrangian}), and we leave this for future work.

It would be useful to examine numerical studies of the square lattice antiferromagnet for logarithmic violations of scaling, Lorentz invariance, and SO(5) symmetry, and compare to our predictions. In particular, we note the violations of scaling observed in Ref.~\cite{Sandvik16}, although for a different square lattice antiferromagnet.

Finally, we note that gapless $\mathbb{Z}_2$ spin liquid studied is an attractive candidate for the ancilla model of doped antiferromagnets \cite{Yahui1,Yahui2,Yahui3}, as it can realize a stable state in the second ancilla layer for the pseudogap state.

As we were completing this paper, we became aware of some related work:\\
({\it i\/}) Superconductivity has been observed \cite{JiangKivelson21,Sheng21} in the doped $J_1$-$J_2$ model; doping the gapless $\mathbb{Z}_2$ spin liquid is a known to be a natural route to $d$-wave superconductivity \cite{SenthilIvanov,SenthilLee05}.\\
({\it ii\/}) Yang {\it et al.\/} \cite{Sandvik21} have detected a gapless spin liquid phase next to the N\'eel phase on the Shastry-Sutherland model, which is obtained from the $J_1$-$J_2$ model by removing 3/4 of the $J_2$ bonds.

\section*{Acknowledgements}

We thank Federico Becca and Anders Sandvik for enlightening discussions which stimulated this work. We are also grateful to Masatoshi Imada, Steve Kivelson, Patrick Ledwith, Ying Ran, T.~Senthil, Ashvin Vishwanath, Chong Wang, and Cenke Xu for valuable discussions. This research was supported by the National Science Foundation under Grant No.~DMR-2002850. 
This work was also supported by the Simons Collaboration on Ultra-Quantum Matter, which is a grant from the Simons Foundation (651440, S.S.).



\appendix

\section{Projective symmetry analysis}
\label{app:psg}

This appendix will present a detailed analysis of the projective symmetry group (PSG) of the Z2A$zz13$ spin liquid, and its neighboring phases. 
Here, we will employ the gauge used by Wen \cite{WenPSG}. Wen described the Z2A$zz13$ spin liquid by the Bogoliubov Hamiltonian in Eq.~(\ref{bogoliubov}) with the ansatz
\bea
u_{\vi,\vi+\hat{x}} &=& \chi \, \tau^x - \eta \, \tau^y \nonumber \\
u_{\vi,\vi+\hat{y}} &=& \chi \, \tau^x + \eta \, \tau^y \nonumber \\
u_{\vi,\vi+\hat{x}+ \hat{y}} &=& - \gamma_1 \, \tau^x \nonumber \\
u_{\vi,\vi-\hat{x}+ \hat{y}} &=&  \gamma_1 \, \tau^x
\label{we1}
\eea
In terms of the
spinons $f_{\vi \alpha}$, this can be written as
\beq
H = - \sum_{\bk} \left[ 2\chi (\cos(k_x) + \cos(k_y)) - i 2\eta (\cos(k_x) - \cos(k_y)) + 4\gamma_1 \sin(k_x) \sin(k_y) \right] f_{-\bk\downarrow} f_{\bk,\uparrow} + \mbox{H.c.}
\eeq 
So in this gauge, the Z2A$zz13$ spin liquid has both $d_{x^2-y^2}+is$ and $d_{xy}$ pairing and no hopping, and the fermion dispersion relation is
\beq
\varepsilon_{\bk}^2 = \left[ 2\chi (\cos(k_x) + \cos(k_y)) + 4\gamma_1 \sin(k_x) \sin(k_y) \right]^2 + \left[2\eta (\cos(k_x) - \cos(k_y))\right]^2 \label{e2}
\eeq
In the ansatz in Eq.~(\ref{we1}), the 3 spin liquids are
\begin{itemize}
    \item The $\pi$-flux phase with SU(2) gauge symmetry corresponds to $\chi=\eta\neq 0$, $\gamma_1=0$.
    \item The `staggered flux' U(1) spin liquid is obtained for $\chi \neq 0$, $\gamma_1 = 0$, $\eta \neq 0$ with $\chi \neq \eta$.
\item The Z2A$zz13$ spin liquid is obtained when the $d_{xy}$ pairing $\gamma_1$ breaks the U(1) down to $\mathbb{Z}_2$. 
\end{itemize}
For our purposes, and in general, a complex Higgs field is needed to break U(1) down to $\mathbb{Z}_2$. We have characterized the $d_{xy}$ pairing above by a real parameter $\gamma_1$, and we need to generalize this to a complex parameter. From the analysis in Section~\ref{sec:gaplessz2}, we deduce that this is obtained by taking a complex $d_{xy}$ order parameter which has opposite phases on the two sublattices {\it i.e.\/}
\bea
u_{\vi,\vi+\hat{x}+\hat{y}} &=&  \left( \begin{array}{cc} 0 & -(\gamma_1 - i \gamma_2) \\
-(\gamma_1 + i \gamma_2)  & 0 \end{array} \right)   \quad, \quad \mbox{$\vi_x + \vi_y$ even} \nonumber \\
u_{\vi,\vi+\hat{x}+\hat{y}} &=&  \left( \begin{array}{cc} 0 & -(\gamma_1 + i \gamma_2) \\
-(\gamma_1 - i \gamma_2)  & 0 \end{array} \right)   \quad, \quad \mbox{$\vi_x + \vi_y$ odd} \nonumber \\
u_{\vi,\vi-\hat{x}+\hat{y}} &=&   \left( \begin{array}{cc} 0 & (\gamma_1 - i \gamma_2)  \\
(\gamma_1 + i \gamma_2) & 0 \end{array} \right)   \quad, \quad \mbox{$\vi_x + \vi_y$ even} \nonumber \\
u_{\vi,\vi-\hat{x}+\hat{y}} &=&   \left( \begin{array}{cc} 0 & (\gamma_1 + i \gamma_2)  \\
(\gamma_1 - i \gamma_2) & 0 \end{array} \right)   \quad, \quad \mbox{$\vi_x + \vi_y$ odd} \,.
\label{we2}
\eea

\subsection{Lattice PSGs}

We first recall the spin liquid classification scheme of Ref.~\onlinecite{WenPSG}.
If $u_{\vi\vj}$ is the mean field ansatz for a spin liquid symmetric under the group action $G$, it transforms as 
\begin{align}
  \mathcal{P}G:\qquad
  u_{\vi\vj} \to W^\dag_g(\vi) u_{G(\vi),G(\vj)} W(\vj) 
\end{align}
where $W_G(\vi)$ is a gauge transform.
In addition to the symmetries, these gauge transformations characterize the spin liquid, yielding the projective symmetry group (PSG) \cite{WenPSG}.

Using the notation from Ref.~\onlinecite{WenPSG}, the spin liquid Z2A$zz13$ is defined by the PSG
\begin{align}\label{eqn:Z2LatPSG}
  W_{tx}(\vi)
  &=
  \tau^0,
  &
  W_{px}(\vi)
  &=
  (-)^{i_x+i_y}i\tau^z,
  &
  W_{pxy}(\vi)
  &=
  i\tau^x,
  \nt
  W_{ty}(\vi)
  &=
  \tau^0,
  &
  W_{py}(\vi)
  &=
  (-1)^{i_x+i_y}i\tau^z,
  &
  W_t(\vi)
  &=
  i\tau^z
\end{align}
while the PSG of U1C$n01n$ (the staggered flux phase) is
\begin{align}\label{eqn:StagFluxLatPSG}
  W_{tx}(\vi)
  &=
  g_3(\theta_x)i\tau^x,
  &
  W_{px}(\vi)
  &=
  (-)^{i_x}g_3(\theta_x)i\tau^x,
  &
  W_{pxy}(\vi)
  &=
  g_3(\theta_{pxy})i\tau^x,
  \nt
  W_{ty}(\vi)
  &=
  g_3(\theta_y)
  i\tau^x,
  &
  W_{py}(\vi)
  &=
  (-)^{i_y}g_3(\theta_y),
  &
  W_t(\vi)
  &=
  (-)^{i_x+i_y}g_3(\theta_t),
\end{align}
where $g_3(\theta)=e^{i\theta \tau^z}$.
From these PSGs we can extract the symmetry fractionalization through the group relations given in the appendix of Ref.~\onlinecite{Thomson17} (Eq.~(B8)).
These are provided in Table~\ref{tab:SymFrac}.
Note that instead of $P_{xy}:(i_x,i_y)\to(i_y,i_x)$, we consider the 90$^\circ$ rotation $R_{\pi/2}=P_{xy}P_y$.
Similarly, $P_x$ is related to the other symmetries through $R_{\pi/2}P_yR_{\pi/2}^{-1}$.

{\renewcommand{\arraystretch}{1.5}
\begin{table}
  \centering
  \begin{tabular}{lclcrcrcr}
    \hline\hline
    & & Group relations & & Z2A$zz13$ & & \multicolumn{1}{c}{U1C$n01n$ lattice}& & \multicolumn{1}{c}{U1C$n01n$ cont}
    \\\hline
    1 &&  $T_y^{-1}T_xT_yT_x^{-1}$  &&  $1$ &&  $e^{-2i(\theta_x-\theta_y)\tau^z}$
    &&  $-e^{-2i(\phi_x-\phi_y)\sigma^z}$
    \\
    2 &&  $P_y^{-1}T_xP_yT_x^{-1}$  &&  $-1$  &&  $e^{-2i\theta_{py}\tau^z}$
    &&  $e^{2i\phi_{py}\sigma^z}$
    \\
    3 &&  $P_y^{-1}T_yP_yT_y$ &&  $-1$  &&  $e^{-2i\theta_{py}\tau^z}$
    &&  $e^{2i\phi_{py}\sigma^z}$
    \\
    4&& $P_y^2$ &&  $-1$  &&  $e^{2i\theta_{py}\tau^z}$
    &&  $e^{-2i\phi_{py}\sigma^z}$
    \\
    5 &&  $P_y^{-1}R_{\pi/2}P_yR_{\pi/2}$ &&  $1$  &&  $-e^{-2i\theta_{py}\tau^z}$
    &&  $-e^{2i\phi_{py}\sigma^z}$
    \\
    6 &&  $R_{\pi/2}^4$ &&  $1$  &&  $\id$
    &&  $\id$
    \\
    7 &&  $R_{\pi/2}^{-1}T_xR_{\pi/2}T_y$ &&  $-1$  &&  $e^{i(2[\theta_{pxy}+\theta_{py}]-[\theta_x+\theta_y])\tau^z}$
    &&  $e^{i(2\phi_r-\phi_x-\phi_y)\sigma^z}$
    \\
    8 &&  $R_{\pi/2}^{-1}T_xR_{\pi/2}T_x^{-1}$ &&  $-1$  &&  $e^{i(2[\theta_{pxy}+\theta_{py}]-[\theta_x+\theta_y])\tau^z}$
    &&  $e^{i(2\phi_r-\phi_x-\phi_y)\sigma^z}$
    \\
    9 &&  $\T^{-1}R_{\pi/2}^{-1}\T R_{\pi/2}$ &&  $-1$  &&  $e^{-2i\theta_t\tau^z}$
    &&  $e^{-2i\phi_t\sigma^z}$
    \\
    10 &&  $\T^{-1}P_y^{-1}\T P_y$ &&  $1$  &&  $\id$
    &&  $\id$
    \\
    11 &&  $\T^{-1}T_x^{-1}\T T_x$ &&  $1$  &&  $-e^{-2i\theta_t\tau^z}$
    &&  $-e^{-2i\phi_t\sigma^z}$
    \\
    12 &&  $\T^{-1}T_y^{-1}\T T_y$ &&  $1$  &&  $-e^{-2i\theta_t\tau^z}$
    &&  $-e^{-2i\phi_t\sigma^z}$
    \\
    13  &&  $\T^2$ &&  $-1$  &&  $e^{2i\theta_t\sigma^z}$
    &&  $e^{2i\phi_t\sigma^z}$
  \end{tabular}
  \caption{Symmetry fractionalization. In keeping with the conventions of Ref.~\onlinecite{Thomson17}, the gauge is chosen such that group relation 7 is fixed to equal $-1$ for the $\mathbb{Z}_2$ spin liquid.}
  \label{tab:SymFrac}
\end{table}
}

\subsection{Identification of staggered flux in continuum model}
The staggered flux state (U1C$n01n$) can be obtained by coupling a Higgs field to the bilinear 
\begin{align}
  \mathcal{O}_3^a&=
  \mathrm{tr}\big(\sigma^a \bar{X}\mu^y(\gamma^xi\partial_y+\gamma^yi\partial_x)X\big)
\end{align}
giving something like
\begin{align}
 \mathcal{L}
  &=
  \mathrm{tr}(\bar{X}\gamma^\mu i\partial_\mu X)
  +
  \Phi_3^a \mathrm{tr}(\sigma^a\bar{X}MX).
\end{align}
The U(1) spin liquid  U1C$n01n$  is then obtained upon condensing one component of $\Phi_3$.
This was determined by considering the symmetry fractionalization of the U(1) spin liquid obtained by condensing the $z$-component of $\Phi_3$: $\langle\Phi_3^z \rangle \neq0$.
Based on the symmetry transformations outlined in Table~\ref{tab:Syms}, this condensate has a corresponding continuum PSG
\begin{align}\label{eqn:StagFluxContPSG}
  V_{tx}&=g_3(\phi_x)i\sigma^x,
  &
  V_{px}&=g_3(\phi_{px}),
  &
  V_{r}&=g_3(\phi_r)i\sigma^x,
  \nt
  V_{ty}&=g_3(\phi_y)i\sigma^x,
  &
  V_{py}&=g_3(\phi_{py}),
  &
  V_t&=g_3(\phi_t),
\end{align}
where $g_3(\phi)=e^{i\phi \sigma^z}$ is an arbitrary gauge transformation. 
Importantly, in the U(1) spin liquid, the phases $\phi$ can take any value.
When these phases are rewritten in terms of the U(1) phases from Eq.~\eqref{eqn:StagFluxLatPSG}, $\theta_G$, according to
\begin{align}
  (\phi_x,\phi_y,\phi_{py},\phi_t,\phi_r)
  &=
  \left(
    \theta_x+\frac{\pi}{4}\cCom
    \theta_y-\frac{\pi}{4}\cCom
    -\theta_y,
    \theta_t,
    \theta_{pxy}+\theta_{py}
  \right)
\end{align}
the symmetry fractionalizations given in columns 4 and 5 of Table~\ref{tab:SymFrac} are identical.
It is possible that two distinct spin liquids (as defined by having distinct PSGs) could nevertheless have identical symmetry fractionalization.
This seems unlikely in this situation and is, moreover, proven false by the explicit derivation of the continuum action from the lattice model.

{\renewcommand{\arraystretch}{1.5}
\begin{table}
  \begin{tabular}{cclccccccccccccr}
    \hline\hline
        && Operators &&  $\T$  &&  $P_x$ &&  $P_y$ &&  $T_x$ &&  $T_y$ && \multicolumn{1}{c}{$R_{\pi/2}$}
      \\\hline
    $\mathcal{O}^a_1$ &&  $\mathrm{tr}\big(\sigma^a\bar{X}\mu^z\gamma^xX\big)$  &&  $-$ &&  $-$ &&  $-$ && $-$ &&  $+$ &&  $-\mathcal{O}^a_2$
    \\
    $\mathcal{O}^a_2$ &&  $\mathrm{tr}\big(\sigma^a\bar{X}\mu^x\gamma^yX\big)$  &&  $-$ &&  $-$ &&  $-$ && $+$ &&  $-$ &&  $-\mathcal{O}^a_1$
    \\
    $\mathcal{O}_3^a$ &&  $\mathrm{tr}\big(\sigma^a \bar{X}\mu^y(\gamma^xi\partial_y+\gamma^yi\partial_x)X\big)$
    &&  $+$ &&  $+$ &&  $+$ &&  $-$ &&  $-$ &&  $-$
    \\
  \end{tabular}
  \caption{}
  \label{tab:Syms}
\end{table}
}

\subsection{Identification of Z2A$zz13$ in continuum model}

The spin liquid Z2A$zz13$ is proximate to U1C$n01n$ in that the PSG of Eq.~\eqref{eqn:Z2LatPSG} may be obtained through gauge transformations and judicious choices of the angles $\theta_G$ in Eq.~\eqref{eqn:StagFluxLatPSG}.
It is, however, simpler to determine the U(1) transformations ({\it i.e.\/} the angles $\phi_\mu$) that map the symmetry fractionalization of U1C$n01n$ to the symmetry fractionalization of Z2A$zz13$.
That is, we find that the assignment
\begin{align}
  (\phi_x,\phi_y,\phi_{py},\phi_t,\phi_r)
  &=
  \left(
    \theta+\frac{\pi}{4}\cCom
    \theta-\frac{\pi}{4} + n_y\pi,
    (2n_{py}+1)\frac{\pi}{2}\cCom
    (2n_{t}+1)\frac{\pi}{2}\cCom
    \theta + (2n_r + n_y + 1)\frac{\pi}{2}
  \right),
  &
  n_\mu\in\mathrm{Z}
\end{align}
transforms the 5th column of Table~\ref{tab:SymFrac} into a set of $\pm1$s that match the third column.
Inserting these $\phi_\mu$s into the PSG defined in Eq.~\eqref{eqn:StagFluxContPSG} and selecting
$n_\mu=0$, $\mu=y,py,t,r$ and $\theta=\pi/4$, we obtain the $\mathbb{Z}_2$ continuum PSG\footnote{The symmetry $P_x$ is related to the other symmetries through $R_{\pi/2}P_yR_{\pi/2}^{-1}$.}
\begin{align}\label{eqn:Z2ContPSG}
  V_{tx}&=-i\sigma^y,
  &
  V_{px}&=\pm i\sigma^z
  &
  V_r&=-\frac{i}{\sqrt{2}}\left(\sigma^x -\sigma^y\right)
  \nt
  V_{ty}&=-i\sigma^x,
  &
  V_{py}&=-i\sigma^z,
  &
  V_t&=
  i\sigma^z.
\end{align}
We can now ask what form of operator needs to couple to a new Higgs field in order to realize this PSG and hence the $\mathbb{Z}_2$ spin liquid Z2A$zz13$.
Firstly, it's clear that the
$\sigma^x$ or $\sigma^y$ components of the Higgs field must condense---condensing in the $\sigma^z$ channel, $\sim \langle{\tilde{\Phi}^z}\rangle\mathrm{tr}(\sigma^z \bar{X}\tilde{M}X)$, would not break the U(1) symmetry.
However, in considering condensates in $x$ or $y$, we see that the 
gauge transformations corresponding to the translations $T_x$ and $T_y$ are different and, further, the rotation $R_{\pi/2}$ exchanges $\sigma^x$ and $\sigma^y$, meaning that both must be present in a symmetric spin liquid.

Based on the symmetry relations documented in Ref.~\onlinecite{Thomson17}, we find that the operators $\mathcal{O}_{1,2}$,
\begin{align}
  \mathcal{O}^a_1&=\mathrm{tr}\big(\sigma^a\bar{X}\mu^z\gamma^xX\big),
  &
  \mathcal{O}^a_2&=\mathrm{tr}\big(\sigma^a\bar{X}\mu^x\gamma^yX\big),
\end{align}
induce the PSG of Eq.~\eqref{eqn:Z2ContPSG} provided they couple to Higgs fields that condense in perpendicular directions.
The symmetry transformation properties of $\mathcal{O}_{1,2}^a$ are given in Table~\ref{tab:Syms}.
That is, given a Lagrangian:
\begin{align}
  \mathcal{L}'
  &=
  \mathrm{tr}(\bar{X}\gamma^\mu i\partial_\mu X)
  +
  {\Phi_1^a}\mathrm{tr}(\sigma^a \bar{X} \mu^z\gamma^x X) 
  +
  {\Phi_2^a}\mathrm{tr}(\sigma^a \bar{X} \mu^x\gamma^y X) 
  + 
  \Phi_3^a\mathrm{tr}\big(\sigma^a \bar{X}\mu^y(\gamma^xi\partial_y+\gamma^yi\partial_x)X\big),
\end{align}
the PSG in Eq.~\eqref{eqn:Z2ContPSG} is obtained when $\langle \Phi_1 \rangle =(\alpha,0,0)$, $\langle \Phi_2 \rangle=(0,\alpha,0)$, and $\langle \Phi_3 \rangle =(0,0,\beta)$, for $\alpha,\beta\in\mathds{R}$.
This agrees with conclusions reached in Section~\ref{sec:MajoranaHiggs}.

\section{Renormalization group analysis of the SU(2) gauge theory}
\label{app:rg}

In this appendix, we describe the origin of the $\log^2$ terms in the critical SU(2) gauge theory in a renormalization group (RG) framework. Integrating the RG equations will lead to an exponentiated prediction for the correlators.

We start with the expression in Eq.~(\ref{eq:selfEnergyLog2}), keep the full Higgs propagator as in Eq.~(\ref{eq:Sigma1}), and perform a standard momentum shell RG in the window $\Lambda - d \Lambda < (p_0^2 + p_x^2 + p_y^2)^{1/2} < \Lambda$
\begin{equation}
\delta \gamma^x \Sigma_1 (k_x) =  \frac{12}{N_f} \int_{\Lambda-d\Lambda}^\Lambda \frac{\dd[3]{p}}{(2\pi)^3} \frac{(p_x + k_x)}{(p_x + k_x)^2 + p_0^2 + p_y^2} 
    \frac{\abs{p}}{p_0^2 + p_y^2 + 4 K \abs{p}p_x^2}\,.
\end{equation}
Expanding to linear order in $k_x$, using spherical co-ordinates with 
\beq
(p_0, p_y, p_x) = \Lambda(\sin\theta \cos \phi, \sin \theta \sin \phi, \cos \theta)\,,
\eeq
and setting $\mu = \cos \theta$, we obtain
\beq
\delta \gamma^x \Sigma_1 (k_x) = \frac{6k_x}{N_f \pi^2} \frac{d \Lambda}{\Lambda} \int_0^1 d \mu (1 - 2 \mu^2)
\frac{1}{1 - \mu^2 + 4 K \Lambda \mu^2}\,.
\eeq
Under normal circumstances, the $\mu$ integral would be a finite numerical constant, and the co-efficient of $d \Lambda/\Lambda$ would the usual RG log which would then contribute (in this case) to the exponent $\eta_\psi$. However, that is not the case here, because of the logarithmic divergence of the $\mu$ integral near $\mu=1$. Evaluating the $\mu$ integral , we obtain
\beq
\delta \gamma^x \Sigma_1 (k_x) = -\frac{3k_x}{N_f \pi^2} \frac{d \Lambda}{\Lambda} \left[ 
\ln \left(\frac{1}{K \Lambda} \right) - 4 + \mathcal{O}(K \Lambda) \right] \,. \label{rg4}
\eeq

In a similar manner, we obtain for the frequency dependence of the self energy
\begin{equation}
\delta \gamma^0 \Sigma_1 (k_0) =  -\frac{12}{N_f} \int_{\Lambda-d\Lambda}^\Lambda \frac{\dd[3]{p}}{(2\pi)^3} \frac{(p_0 + k_0)}{(p_0 + k_0)^2 + p_x^2 + p_y^2} 
    \frac{\abs{p}}{p_0^2 + p_y^2 + 4 K \abs{p}p_x^2}\,.
\end{equation}
In spherical co-ordinates this simplifies to 
\bea
\delta \gamma^0 \Sigma_1 (k_0) &=& -\frac{6k_0}{N_f \pi^2} \frac{d \Lambda}{\Lambda} \int_0^1 d \mu (1 -  (1-\mu^2))
\frac{1}{1 - \mu^2 + 4 K \Lambda \mu^2} \nonumber \\
&=& -\frac{3k_0}{N_f \pi^2} \frac{d \Lambda}{\Lambda} \left[ 
\ln \left(\frac{1}{K \Lambda} \right) - 2 + \mathcal{O}(K \Lambda) \right]
\,. \label{rg5}
\eea
The expression for $\delta \gamma^y \Sigma_1 (k_y)$ is the same as $\delta \gamma^0 \Sigma_1 (k_0)$, after mapping $k_0 \Rightarrow k_y$.

We can also examine the vertex correction for the SO(5) order parameter in a similar manner. From Eq.~(\ref{eq:so5logSquaredIsolation}) at zero external momentum, we note that the vertex correction needs the integral
\bea
\delta V &=& \frac{1}{N_f} \int_{\Lambda-d\Lambda}^\Lambda \frac{\dd[3]{p}}{(2\pi)^3} \frac{1}{p_x^2 + p_y^2 + p_0^2} \frac{4\abs{p}}{p_0^2 + p_y^2+4 K \abs{p}p_x^2} \nonumber \\ &=& \frac{1}{2 N_f \pi^2} \frac{d\Lambda}{\Lambda} \int_{0}^1 d \mu \frac{4}{1-\mu^2 + 4 K \Lambda \mu^2}\nonumber \\
&=& \frac{1}{N_f \pi^2} \frac{d \Lambda}{\Lambda} \left[ 
\ln \left(\frac{1}{K \Lambda} \right) + \mathcal{O}(K \Lambda) \right]
\,. \label{eq:deltaV}
\eea

We now proceed as usual to obtain the RG equations from the momentum shell results under the rescaling
\bea
x' &=& x e^{-\ell} \nonumber \\
y' &=& y e^{-\ell}. \nonumber \\
\tau' &=& \tau \exp \left( - \int_0^\ell d\ell' z(\ell') \right) \label{rg1}
\eea
Importantly, we note the flow of the irrelevant coupling $K$ under this transformation
\beq
\frac{dK}{d \ell} = - K\,. \label{rg2}
\eeq
For the fermion field we define
\beq
\psi' = \psi \, \exp \left( \int_0^\ell d\ell' \, \frac{1+z(\ell')+ \eta_\psi (\ell')}{2} \right) \label{rg3}
\eeq
The field $\psi$ is not gauge-invariant, and neither is its anomalous dimension $\eta_\psi$. However, the leading $\log^2$ term we shall find shortly is gauge invariant. In the presence of the $\log^2$ term, we will also see that the usual logarithm terms have a non-universal co-efficient. So we ignore the gauge field contributions here (the gauge field induced renormalizations have been computed in Refs.~\cite{RanWen06,YingRanThesis}), because they only contribute logarithm terms which become part of overall terms which are non-universal. 

Matching Eqs.~(\ref{rg4},\ref{rg5}) to Eqs.~(\ref{rg1},\ref{rg3}) we obtain
\bea
\eta (\ell) &=& \frac{6}{N_f \pi^2} \left[ \ln \left(\frac{1}{K(\ell) \Lambda} \right) -3 \right] \nonumber \\
z(\ell) &=& 1 + \frac{6}{N_f \pi^2} \label{rg6}
\eea
Assuming a bare value $K(0)=K_0$, integrating Eq.~(\ref{rg2}) to obtain $K(\ell) = K_0 e^{- \ell}$, and then integrating Eq.~(\ref{rg6}) we obtain
\beq
\int_0^\ell \eta(\ell') d \ell' = \frac{6}{N_f \pi^2} \left[ \frac{\ell^2}{2} - \left( \ln(K_0 \Lambda) + 3 \right) \ell \right] \label{rg7}
\eeq
We can now obtain the momentum dependence of physical observables by evaluating them at a scale $\ell = \ell^\ast = \ln(\Lambda/|p|)$.  
Note that the co-efficient of $\ell^\ast$ involves the bare value of $K_0$, and hence the co-efficient of the logarithm term is non-universal, as claimed earlier. 
The leading term is $\log^2$, and its co-efficient is universal and agrees with that in Eq.~(\ref{eq:Zpsi}); similarly, Eq.~(\ref{eq:deltaV}) agrees with Eq.~(\ref{eq:Zso5}). Inserting the integral Eq.~(\ref{rg7}) into Eq.~(\ref{rg3}), we obtain results of the form in Eq.~\eqref{chirg}.

\section{Isolation of logarithm-squared divergences in one-loop corrections}
\label{app:oneLoopCalcs}
We state in the main text that logarithm-squared divergences in the critical SU(2) gauge theory arise in the one-loop diagrams in a certain parameter regime, given by Eq.~\eqref{eq:logSquaredLimit}. This is shown in the main text for the simplest one-loop calculation, which is the $\Phi_1$ ($\Phi_2$) contribution to the fermion self-energy with external momenta $k_x$ ($k_y$). Here, we provide more general calculations for other cases.

We first analyze the $\Phi_1$ contribution to the fermion self-energy with external momenta $k_0$. This is equivalent to the $k_y$ external momenta, as well as the $\Phi_2$ contribution with external momenta $k_0$, $k_x$.
\begin{equation}
  \begin{aligned}
    \gamma^0 \Sigma(k_0) &\approx - \frac{12}{N_f} \int \frac{\dd[3]{p}}{(2\pi)^3} \frac{p_0 + k_0}{(p_0+k_0)^2 + p_x^2 + p_y^2} \frac{\abs{p_x}}{p_0^2 + p_y^2 + 4 K \abs{p_x}^3 }
    \\
     &=-\frac{12}{N_f} \int \frac{\dd{p_x}}{8\pi^2} \frac{\abs{p_x} (p_x^2 - 4K \abs{p_x}^3 - k_0^2)}{k_0 \sqrt{-(4K)^2 \abs{p_x}^6 + 8K \abs{p_x}^3 (p_x^2 - k_0^2) - (p_x^2 + k_0^2)^2} } 
    \\
    &\Bigg[ \tan^{-1} \left( \frac{-p_x^2 + 4K \abs{p_x}^3 - k_0^2 }{\sqrt{-4K^2 \abs{p_x}^6 + 8 K \abs{p_x}^3 (p_x^2 - k_0^2) - (p_x^2 + k_0^2)^2}}  \right) 
    \\
  &+ \tan^{-1} \left( \frac{-p_x^2 + 4 K \abs{p_x}^3 + k_0^2 }{\sqrt{-(4K)^2 \abs{p_x}^6 + 8 K \abs{p_x}^3 (p_x^2 - k_0^2) - (p_x^2 + k_0^2)^2}}  \right) \Bigg]
 \\
 &- \frac{12}{N_f} \int \frac{\dd{p_x}}{8\pi^2} \frac{\abs{p_x}}{2 k_0} \ln\left( \frac{p_x^2}{4 K \abs{p_x}^3} \right)\,.
 \label{eq:selfEnergyFreqLog2}
  \end{aligned}
\end{equation}
We see that the dominant term is proportional to $k_0 \log^2(K k_0)$, arising in the same limit as in Eq.~\ref{eq:logSquaredLimit}. This $\log^2$ dependence comes from the inverse tangents, since $2 \tan^{-1}(x) = i \log\left( \frac{1+ix}{1-ix} \right)$. Assuming $K \abs{p_x}^3 \ll p_x^2$, the expression in the denominator of the inverse tangent argument is $\approx i p_x^2$, so our integrand $\approx i \tan^{-1} \left( i (1 + 4K \abs{p_x}) \right)$. If we further assume $K \abs{p_x} \ll 1$, we get an integrand that scales like $k_0/p_x \ln(K p_x)$, and hence the full expression scales as $k_0 \log^2 (K k_0)$.

\begin{figure}
  \begin{center}
    \includegraphics[width=5in]{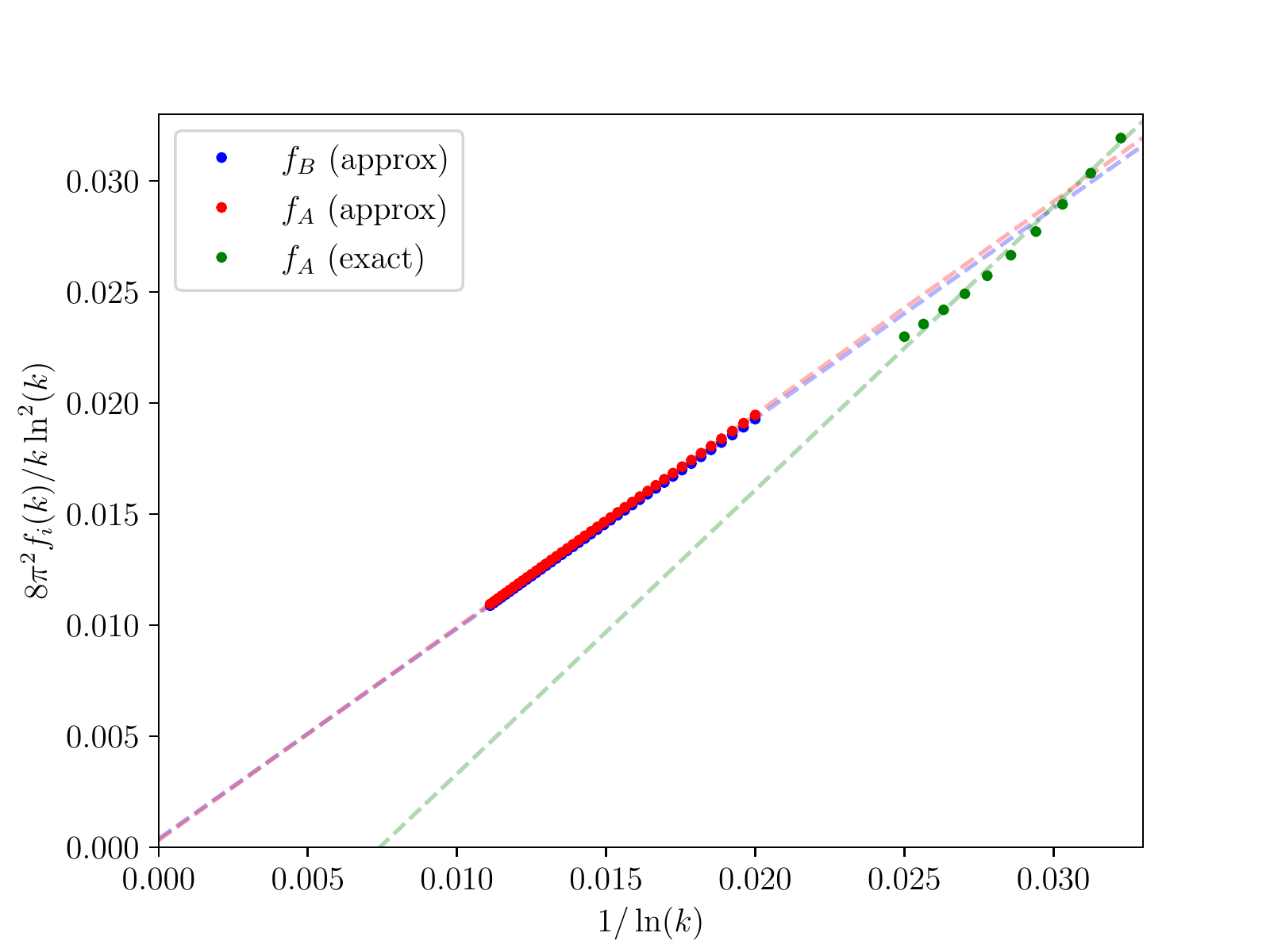}
\end{center}
\caption{Denoting the integrands of the two types of self-energy contributions in Eq.~\eqref{eq:selfEnergyLog2} and Eq.~\eqref{eq:selfEnergyFreqLog2} with $K=1$ as $f_A(k_x)$ and $f_B(k_0)$, we plot a numerical evaluation of $8 \pi^2 f_i(k)/(k \ln^3(k))$ vs ${1}/({\ln(k)})$. The form of this expression is designed to isolate the $\log^2$ contribution at small momenta, and agreement with our analytic predictions should give a straight line with a slope of $1$. The approximation of the Higgs propagator as Eq.~\ref{eq:higgsApprox} allows for greater numerical precision, as the dimensionality of the integral can be reduced by performing portions of the integral analytically. These numerical evaluations give good agreement with analytic predictions as well as calculations using the full form of the Higgs propagator.}
\label{fig:numericalIntegrations}
\end{figure}

These $\log^2$ contributions are verified by numerically evaluating Eq.~\ref{eq:selfEnergyLog2} and Eq.~\ref{eq:selfEnergyFreqLog2} and analyzing the behavior at small $k$, as shown in Fig.~\ref{fig:numericalIntegrations}.

For the Higgs vertex correction to the $\SO(5)$ order parameter, we can also isolate a $\log^2$ divergence.
Calculating the $\Phi_1$ correction to the vertex, we regulate the integral by including an external momenta $2k_x$ evenly distributed between the two outgoing fermions. The vertex correction is
\begin{equation}
  \begin{aligned}
   & \frac{\mu^z \sigma^a \Gamma^i \sigma^a \mu^z}{N_f}\int \frac{\dd[3]{p}}{(2\pi)^3} \frac{p_x^2 - k_x^2 + p_y^2 + p_0^2}{\left[ (p_x + k_x)^2 + p_y^2 + p_0^2 \right] \left[ (p_x - k_x)^2 + p_y^2 + p_0^2 \right]} \frac{4 \abs{p_x}}{p_0^2 + p_y^2 + 4 K \abs{p_x}^3}
   \\
   &= \frac{\mu^z \sigma^a \Gamma^i \sigma^a \mu^z}{N_f}\int \frac{r \dd{r} \dd{p_x}}{(2\pi)^2} \frac{p_x^2 - k_x^2 + r^2}{\left[ (p_x + k_x)^2 + r^2 \right]\left[ (p_x - k_x)^2 + r^2 \right]} \frac{4\abs{p_x}}{r^2 + 4K \abs{p_x}^2}
   \\
   &= \frac{\mu^z \sigma^a \Gamma^i \sigma^a \mu^z K}{N_f} \int \frac{\dd{p_x}}{4\pi^2} \frac{\abs{p_x} }{k_x p_x \left[ (p_x - x_x)^2 - 4 K \abs{p_x}^2 \right] \left[ (p_x + k_x)^2 - 4 K \abs{p_x}^2 \right]}
   \\
   &\times \Bigg[ \left( (p_x^2 - k_x^2) \abs{p_x}^2 + \frac{(p_x + k_x)^2 (p_x - k_x)^2}{4K} \right) \ln \left( \frac{(p_x + k_x)^2}{(p_x - k_x)^2} \right)
   \\
 &+ (p_x + k_x)^2 \left[ \abs{p_x}^3 \ln \left( \frac{4K \abs{p_x}^3}{(p_x + k_x)^2} \right) + \frac{(k_x^2 - p_x^2)}{4 K } \ln \left( \frac{4K \abs{p_x}^3}{(p_x - k_x)^2} \right)\right]
\\
 &- (p_x - k_x)^2 \left[ \abs{p_x}^3 \ln \left( \frac{4K \abs{p_x}^3}{(p_x - k_x)^2} \right) + \frac{(k_x^2 - p_x^2)}{4K} \ln \left( \frac{4K \abs{p_x}^3}{(p_x + k_x)^2} \right)\right]
 \Bigg]
 \label{eq:so5logSquared}
  \end{aligned}
\end{equation}
We can obtain a $\log^2$ from the second logarithm in the brackets in the limit given by Eq.~\eqref{eq:logSquaredLimit}.

\section{Alternate computation of $z$ in the SU(2) gauge theory}
\label{app:dimreg}
In the main text we emphasize that although irrelevant terms in the Higgs propagator turn out to strongly influence the renormalization of the critical theory, these effects cancel in the dynamical critical exponent, and its value can be computed through more standard methods. In this section, we compute $z$ via dimensional regularization after explicitly setting the irrelevant Higgs terms to zero, and show that is gives the same answer for $z$ as in the main text. For the calculation of $z$, we are interested in the counterterms generated by 
\begin{equation}
  \begin{aligned}
    \gamma^0 \pdv{\Sigma}{k_0}\Big|_{k=0} - \gamma^x \pdv{\Sigma}{k_x}\Big|_{k=0} &= -\frac{6}{N_f} \int \frac{\dd[3]{p}}{(2\pi)^3}\left[ \frac{p_y^2  }{p^4 \Gamma_1(p)}  -  \frac{p_0^2 - p_x^2}{p^4 \Gamma_2(p)}\right]
    \label{eq:selfEnergyDifference}
  \end{aligned}
\end{equation}
This integrand is well-behaved for $p^2 \neq 0, \infty$, and one can see that the second term in brackets vanishes, since the $\Gamma_2$ propagator is invariant under $p_x \leftrightarrow p_0$.

  The integrals over $p_0, p_y$ can be performed exactly in radial coordinates, which gives
\begin{equation}
  \begin{aligned}
    - \frac{3}{\pi^2 N_f} \int_{-\infty}^\infty \dd{p_x} \frac{1}{\abs{p_x}} = - \frac{6}{\pi^2 N_f} \int_0^\infty \frac{\dd{p_x}}{p_x}\,.
  \end{aligned}
\end{equation}
Continuing the $p_x$ integral to $1-\epsilon$ dimensions and imposing a UV cutoff $\Lambda$, yields
\begin{equation}
  \begin{aligned}
    -\frac{6}{\pi^2 N_f} \mu^{\epsilon} \int_0^\Lambda \frac{\dd{p_0}}{p_0^{1+\epsilon}} = -\frac{6}{\pi^2 N_f\epsilon} \mu^{\epsilon} \Lambda^{-\epsilon} = -\frac{6}{\pi^2 N_f} \left( \frac{1}{\epsilon} + \ln \left( \frac{\mu}{\Lambda} \right) + \order{\epsilon}\right)
  \end{aligned}
\end{equation}
Which gives the same answer for $z$ as when the irrelevant Higgs terms were used to regulate the divergences in the self-energy.

\section{Higgs field renormalization in the SU(2) gauge theory}

For completeness, we compute the $\log^2$ corrections in the critical SU(2) gauge theory to the Yukawa couplings at one-loop level, since these determine the renormalization of the Higgs fields. The calculations are nearly identical to those of the $\SO(5)$ order parameter.

The correction to the $\Phi_1$ Yukawa coupling is given by the integral
\begin{equation}
  \begin{aligned}
    &\frac{(\mu^z  \sigma^b)(\mu^z  \sigma^a)(\mu^z  \sigma^b)}{N_f} \int \frac{\dd[3]{p}}{(2\pi)^3}
    \gamma^x \frac{\slashed{p} - \slashed{k_1}}{(p - k_1)^2} \gamma^x \frac{\slashed{p} - \slashed{k_2}}{(p - k_2)^2} \gamma^x \frac{1}{\Gamma_1(p) + K p_x^2}
\\
    +&\frac{(\mu^x  \sigma^b)(\mu^z  \sigma^a)(\mu^x  \sigma^b)}{N_f} \int \frac{\dd[3]{p}}{(2\pi)^3}
    \gamma^y \frac{\slashed{p} - \slashed{k_1}}{(p - k_1)^2} \gamma^x \frac{\slashed{p} - \slashed{k_2}}{(p - k_2)^2} \gamma^y \frac{1}{\Gamma_2(p) + K p_y^2}
    \label{eq:yukawaOneLoop}
  \end{aligned}
\end{equation}
Evaluating the first term in the limit in Eq.~\ref{eq:logSquaredLimit}, we set the external momenta to zero and use it as an IR cutoff $k$, which gives to $\log^2$ order
\begin{equation}
  \begin{aligned}
    &\frac{(\mu^z  \sigma^b)(\mu^z  \sigma^a)(\mu^z  \sigma^b)}{2 \pi^2 N_f} \ln^2(K k) \,.
  \end{aligned}
\end{equation}
This coefficient is identical to the $\SO(5)$ correction, as the two integrals are the same to leading order in $k$. The contribution is the same for the second term in Eq.~\eqref{eq:yukawaOneLoop}, giving the final Yukawa correction
\begin{equation}
    -\frac{\mu_z \sigma^a}{\pi^2 N_f}\ln^2(K k)\,. \label{appd1}
\end{equation}
The correction to the $\Phi_2$ Yukawa term is identical, as the two are related by a spatial rotation. 

We now renormalize the $\Phi_{1,2}$ fields so that the Yukawa coupling remains invariant, as in Ref.~\cite{Huh08}.
Hence, the Higgs fields are renormalized at $\log^2$ order, $\Phi_i = \sqrt{Z_\Phi} \Phi_{i, R}$, with corrections from $Z_\psi$ and (\ref{appd1})
\begin{equation}
    Z_\Phi = 1 + \frac{6}{\pi^2 N_f} \ln^2(K \mu) +\frac{2}{\pi^2 N_f}\ln^2(K \mu) = 1 + \frac{8}{\pi^2 N_f} \ln^2(K \mu)
\end{equation}
\section{Evaluation of two-loop SO(5) order parameter corrections}
\label{app:twoLoop}
\begin{figure}
    \centering
    \includegraphics[width=5in]{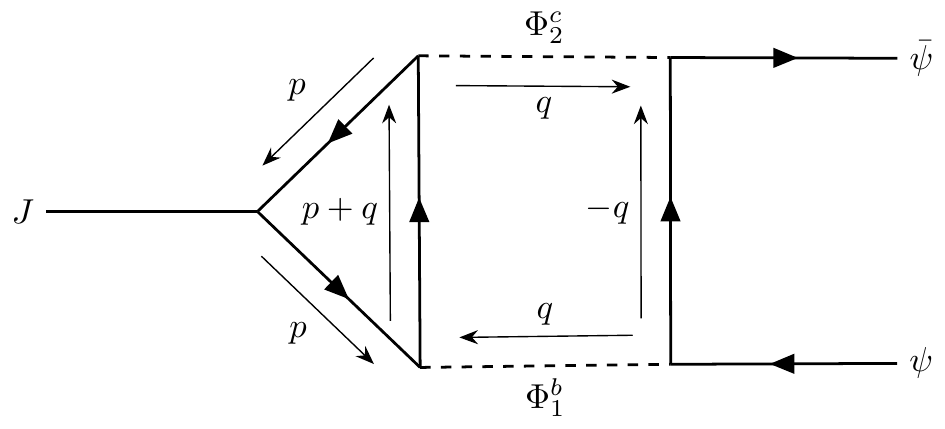}
    \caption{The $\mathcal{O}(N_f^{-1})$ two-loop correction to the SO(5) order parameter. We set all external momenta to zero. Shown is one of four possible diagrams - the other three can be obtained by either exchanging $\Phi_1 \leftrightarrow \Phi_2$, crossing the lines of the Higgs propagators, or both. All give the same contribution at zero external momenta.}
    \label{fig:twoLoop}
\end{figure}

In this appendix, we evaluate the $\order{N_f^{-1}}$ two-loop correction to the SO(5) order parameter, shown in the main text and displayed here in Fig.~\ref{fig:twoLoop} with internal momenta labeled. The diagram shown is one of four possible contributions - additional diagrams can be generated by either exchanging $\Phi^1 \leftrightarrow \Phi^2$ or crossing the propagators of the Higgs bosons, but all give the same correction for zero external momenta. The main conclusion of this appendix is that this contribution is well-behaved upon setting the dangerously irrelevant operators to zero and only contributes standard logarithm divergences, which we argue in the main text and in Appendix~\ref{app:rg} give non-universal corrections to the order parameter scaling. Intuitively, this may be thought of as related to the fact that these two-loop diagrams require \textit{both} types of Higgs $\Phi_{1,2}$, as they vanish trivially when both Higgs propagators are of the same type. As the $\log^2$ divergences are connected to the rotational symmetry breaking in the $\order{N_f}$ effective action for the Higgs propagators, it is natural - although still a non-trivial fact - that these two-loop diagrams which respect rotational symmetry only contribute single logarithm divergences. 

This two-loop correction vanishes for the VBS order parameter, so we focus on the N\'eel order parameter, where the source vertex contributes a factor of $\mu^y \sigma^a$. 
We first evaluate the fermion loop integral,
\begin{equation}
\begin{aligned}
     &(-1) N_f \Tr  \sigma^b \sigma^c \mu^y \mu^x \mu^z \int \frac{\dd[3]{p}}{(2\pi)^3} \frac{\slashed{p}}{p^2} \gamma^x \frac{\slashed{p} + \slashed{q}}{(p+q)^2} \gamma^y \frac{\slashed{p}}{p^2}
      \\
      &= -4 N_f \delta_{bc} \Tr\left[ \gamma^\mu \gamma^x \gamma^\nu \gamma^y \gamma^\sigma \right] \int \frac{\dd[3]{p}}{(2\pi)^3} \frac{p_\mu (p+q)_\nu p_\sigma}{p^4 (p+q)^2}
      \\
      &= - 8i N_f \delta_{bc} \left( \delta^{\mu x} \epsilon^{y \nu \sigma} + \delta^{\nu y} \epsilon^{x \sigma\mu} + \delta^{\sigma y} \epsilon^{x \nu \mu} - \delta^{\nu \sigma} \delta^{\mu 0} \right)\int \frac{\dd[3]{p}}{(2\pi)^3} \frac{p_\mu (p+q)_\nu p_\sigma}{p^4 (p+q)^2}\,.
      \label{eq:fermionLoop}
      \end{aligned}
\end{equation}
The integral over $p$ yields
\begin{equation}
   \int \frac{\dd[3]{p}}{(2\pi)^3} \frac{p_\mu (p+q)_\nu p_\sigma}{p^4 (p+q)^2} =  \frac{1}{128} \frac{1}{\abs{q}} \left[ 3 \delta^{\mu\sigma} q_\nu - \delta^{\sigma\nu} q_\mu - \delta^{\mu\nu} q_\sigma + \frac{q_\mu q_\nu q_\sigma}{q^2} \right]
\end{equation}
and contracting with the tensors in Eq.~\eqref{eq:fermionLoop} gives the final contribution of the fermion loop
\begin{equation}
    \frac{iN_f}{2} \frac{q_0}{\abs{q}} \delta_{bc} \,.
\end{equation}
We combine this with the remaining loop integral, setting the coefficient $K$ of the irrelevant operators to zero, to give
\begin{equation}
    \begin{aligned}
      &\frac{i}{2N_f} \delta_{bc} \mu^z \mu^x \sigma^b \sigma^c \int \frac{\dd[3]{q}}{(2\pi)^3}  \frac{q_0}{\abs{q}} \gamma^x \frac{(-\slashed{q})}{q^2} \gamma^y \frac{16 q^2}{(q_0^2 + q_x^2)(q_0^2 + q_y^2)} \\ 
      &=-\frac{16  \mu^y \sigma^a}{N_f}  \int \frac{\dd[3]{q}}{(2\pi)^3} \frac{q_0^2}{\abs{q}}\frac{1}{(q_0^2 + q_x^2)(q_0^2 + q_y^2)}\,.
    \end{aligned}
  \end{equation}
  Focusing on the integrand, we can compute this by converting to radial coordinates,
  \begin{equation}
    \begin{aligned}
      &\int \frac{\dd{z} \dd{\theta} r \dd{r}}{(2\pi)^3} \frac{z^2}{\sqrt{z^2 + r^2}} \frac{1}{(z^2 + r^2 \cos^2\theta)(z^2 + r^2 \sin^2\theta)}
      \\
      &= \frac{1}{2 \pi^2} \int \dd{z} \dd{r} \frac{\abs{z} r}{(z^2 + r^2) ( 2 z^2 + r^2)} 
      \\
      &= \frac{\ln 2}{4\pi^2} \int \dd{z} \frac{1}{\abs{z}}\,.
    \end{aligned}
  \end{equation}
  Hence, this two-loop contribution only contributes a standard logarithm divergence, and is subleading in comparison to the one-loop Higgs corrections.
  
  We also analyze the two-loop corrections to the $\bar{\psi} \psi$ bilinear, whose symmetry properties correspond to the scalar spin chirality. This is motivated by the fact that $\log^2$ terms in the $\order{1/N_f}$ one-loop corrections exactly cancel the $\log^2$ self-energy terms. Hence, if two-loop corrections only contributed standard logarithm divergences, then the scalar spin chirality would have a power law decay at $\order{1/N_f}$. In fact, the two-loop corrections involving two Higgs propagators vanish exactly. If the Higgs propagators are different, as was the case for the N\'eel corrections, then the trace over $\mu$ in the fermion loop vanishes. If the Higgs propagators are the same, then the trace over $\gamma$ vanishes, since
  \begin{equation}
      \Tr[\gamma^\mu \gamma^x \gamma^\nu \gamma^x \gamma^\sigma] p_\mu (p+q)_\nu p_\sigma = \Tr[\gamma^\mu \gamma^y \gamma^\nu \gamma^y \gamma^\sigma] p_\mu (p+q)_\nu p_\sigma = 0
  \end{equation}

\bibliography{z2neel}

\end{document}